\documentclass[pdflatex,sn-vancouver-num]{sn-jnl}

\usepackage{blkarray}
\usepackage{graphicx}%
\usepackage{multirow}%
\usepackage{amsmath,amssymb,amsfonts}%
\usepackage{amsthm}%
\usepackage{mathrsfs}%
\usepackage[title]{appendix}%
\usepackage{xcolor}%
\usepackage{textcomp}%
\usepackage{manyfoot}%
\usepackage{booktabs}%
\usepackage{longtable}%
\usepackage{algorithm}%
\usepackage{algorithmicx}%
\usepackage{algpseudocode}%
\usepackage{listings}%

\theoremstyle{thmstyleone}%
\theoremstyle{thmstyletwo}%

\theoremstyle{thmstylethree}%
\newcommand{\argmin}[1]{\underset{#1}{\text{argmin}}}

\usepackage{geometry}

\begin{document}

\title[Higher-Order Network Structure Inference]{Higher-Order Network Structure Inference: A Topological Approach to Network Selection}


\author*[1]{\fnm{Adam} \sur{Schroeder}}\email{aschr@g.ucla.edu}

\author[3]{\fnm{Russell J.} \sur{Funk}}\email{rfunk@umn.edu}

\author[4]{\fnm{Jingyi} \sur{Guan}}\email{gjingyi@uw.edu}

\author[2]{\fnm{Taylor}
\sur{Okonek}}\email{tokonek@macalester.edu}

\author[2]{\fnm{Lori}
\sur{Ziegelmeier}}\email{lziegel1@macalester.edu}

\affil[1]{\orgdiv{Department of Mechanical and Aerospace Engineering}, \orgname{University of California, Los Angeles}, \orgaddress{\street{405 Hilgard Avenue}, \city{Los Angeles}, \postcode{90095}, \state{CA}, \country{USA}}}

\affil[2]{\orgdiv{Department of Mathematics, Statistics, and Computer Science}, \orgname{Macalester College}, \orgaddress{\street{1600 Grand Avenue}, \city{Saint Paul}, \postcode{55105}, \state{MN}, \country{USA}}}

\affil[3]{\orgdiv{Carlson School of Management}, \orgname{University of Minnesota}, \orgaddress{\street{321 19th Avenue South}, \city{Minneapolis}, \postcode{55455}, \state{MN}, \country{USA}}}

\affil[4]{\orgdiv{Department of Biostatistics}, \orgname{University of Washington}, \orgaddress{\street{3980 15th Avenue NE}, \city{Seattle}, \postcode{98195}, \state{WA}, \country{USA}}}


\abstract{Thresholding---the pruning of nodes or edges based on their properties or weights---is an essential preprocessing tool for extracting interpretable structure from complex network data, yet existing methods face several key limitations. Threshold selection often relies on heuristic methods or trial and error due to large parameter spaces and unclear optimization criteria, leading to sensitivity where small parameter variations produce significant changes in network structure. Moreover, most approaches focus on pairwise relationships between nodes, overlooking critical higher-order interactions involving three or more nodes. We introduce a systematic thresholding algorithm that uses topological data analysis to identify optimal network parameters by accounting for higher-order structural relationships. Our method uses persistent homology to favor thresholds at which the network's topological representation is stable under small parameter changes, subject to minimum requirements for higher-order features. Hyperparameters allow users to specify minimum requirements for topological features, effectively constraining the parameter search to avoid spurious solutions. We demonstrate the approach using networks of concepts linked by co-occurrence in scientific abstracts. Across scientific fields, paired comparisons with ten alternative thresholds favor our method at the median for finite-feature density and normalized total persistence in homological dimensions one and two. Our approach allows researchers to incorporate domain-specific constraints and extends beyond network thresholding to general parameterization problems in data analysis.}




\maketitle
\section{Introduction}\label{sec:introduction}

The growing availability of large-scale relational data---from social media interactions \citep{cinelli2021echo, rajkumar2022causal, park2018strength} and biological systems \citep{guimaraes2020structure, giusti2015clique, bullmore2009complex} to human mobility patterns \citep{louail2015uncovering, schlosser2020covid, riascos2020networks} and semantic networks \citep{rule2015lexical, hofstra2020diversity, christianson2020architecture}---offers unprecedented opportunities for the study of complex networks. Yet this abundance is not without challenges. The data are often noisy, and their volume makes identifying and characterizing a network’s most important properties---and uncovering meaningful underlying structures---difficult, whether through quantitative methods or qualitative approaches (e.g., visualization) \citep{padgett2020political, powell2005network}. 

Within this context, researchers have devoted increasing attention to developing and applying methods for ``pruning’’ or ``thresholding’’ network data to yield more meaningful and analytically tractable structures \citep{de2010inferring, kossinets2009origins}. Sometimes referred to as ``backbone'' identification \citep{dai2018transport, neal2022backbone, serrano2009extracting, domagalski2021backbone, neal2014backbone}  or ``network structure inference'' \citep{brugere2018network}, these techniques typically serve as a preprocessing step in the network analysis pipeline. Given a graph $G = (V, E)$, thresholding methods often assign real-valued weights to edges (E), denoted as $\omega: E \rightarrow \mathbb{R}$, representing properties like the frequency of interaction, although nodes (V) may also be assigned weights. These weights are then used to determine the inclusion or exclusion of nodes or edges in the network under consideration based on a specified threshold. 

While existing approaches for network thresholding are valuable, they suffer from several important limitations. First, selection of appropriate thresholds is itself a challenging problem due to the large parameter space (i.e., potential features on which to threshold), the granularity of node and edge attributes, the lack of reference or ``ground truth'' networks, and unclear optimization criteria \citep{brugere2018network}. Consequently, thresholds are often chosen using trial and error or heuristic methods. Moreover, variations on $\omega$ may lead to non-ignorable changes in the resulting network, implying a significant degree of sensitivity to the precise value used \cite{de2010inferring}.

Second, existing approaches are limited by their focus on \emph{lower-order} network structures, primarily node properties or dyadic interactions (edges). Importantly, network structures involving \emph{higher-order} interactions, encompassing groups of three or more nodes, are not only common in many real-world networks but also increasingly recognized as pivotal to network structure and dynamics \citep{benson2016higher, lambiotte2019networks, bick2023higher, battiston2020networks, zapata2022invitation}. By making thresholding decisions based on dyadic interactions or node characteristics, existing methods risk overlooking the importance of seemingly less significant nodes or edges that may be nevertheless crucial for the large-scale architecture of the network. This narrow focus could lead to an inaccurate representation of the network's true structure.

In this study, we address these limitations by drawing on techniques from topological data analysis \citep{carlsson2021topological, dey2022computational}. Our method encodes interactions among nodes as $n$-dimensional simplices, enabling the incorporation of information on higher-order network structure. Nodes are encoded as 0-dimensional simplices, edges as 1-dimensional, triangles (i.e., groups of three nodes) as 2-dimensional, tetrahedra (i.e., groups of four nodes) as 3-dimensional, and so on. Our method therefore captures both basic node and edge information (as is done in existing approaches) and more complex structures up to a dimension $k$. This dimension is chosen based on the analyst's understanding of the substantive context and computational resources. Our method incorporates higher-order structure in two senses---simplices represent groups of three or more pairwise-connected nodes, and persistent homology characterizes global network organization through cycles and voids. These topological features depend on how the simplices connect and fill, beyond what counts of nodes, edges, and simplices alone reveal. In addition, our approach supports different thresholding types (e.g., node- or edge-based) and allows the analyst to apply multiple thresholds or thresholding criteria to the same network.

For a given network or related data structure (e.g., a correlation matrix), we use persistent homology to analyze a range of potential thresholds across the parameter space. We optimize to determine the thresholds that favor stability of the network's topological representation under small parameter changes. The algorithm minimizes a local change objective based on distances between persistence image vectors at neighboring parameter choices, subject to minimum feature counts specified through dimension-specific hyperparameters.

We demonstrate the method on networks of concepts extracted from scientific texts, a common object of study in the Science of Science. Nodes represent concepts extracted from research abstracts, and edges connect concepts that co-appear in a publication. We use the field of applied mathematics as a detailed example. These networks describe relationships among concepts within a field, but noise in the extracted terms can obscure meaningful patterns. The thresholding problem is to select frequency bounds that retain meaningful relationships without excluding less common concepts relevant to emerging or specialized research areas. 

We compare topological thresholding with ten alternatives based on fixed node-frequency cutoffs, global edge-weight cutoffs, and statistical backbone thresholds. The comparison measures the density and normalized total persistence of finite one- and two-dimensional homological features, first in the applied mathematics example and then across a variety of scientific fields. In the applied mathematics example, topological thresholding exceeds all ten alternatives on all four measures; across scientific fields, the median paired comparison favors topological thresholding over each alternative.

The remainder of the paper is organized as follows. In Section \ref{sec:background}, we provide mathematical background on persistent homology and persistence images, and discuss how we apply these methods to concept networks. In Section \ref{sec:algorithm}, we present the algorithm and discuss its key features. In Section \ref{sec:application}, we apply the algorithm to concept networks, focusing on the field of applied mathematics as a case study. In Section \ref{sec:statistical_implications}, we examine how the network selected for local stability compares with topological variation across the threshold grid. Section~\ref{subsec:thresholding_comparison} compares the method with alternative thresholds in the applied mathematics example and across scientific fields. Finally, Section \ref{sec:discussion} offers concluding remarks.


\section{Background}\label{sec:background}
In this section, we provide the mathematical background needed to apply topological data analysis to network data. We introduce persistent homology and persistence images, the key tools that enable our approach. We conclude by describing the specific challenge of thresholding concept networks that motivates our application in Section \ref{sec:application}.

\subsection{Persistent Homology}\label{subsec:persistent_homology}

\emph{Homology}, a core mathematical concept within algebraic topology, was initially formulated to classify topological spaces by examining their invariant structures \citep{hatcher2002algebraic}.
These structures, often intuited as `holes' in specific dimensions of the space, represent features like connected components, loops, trapped volumes, and so on in dimensions $k= 0, 1, 2$, and beyond, respectively. Details on computing homology groups are given in Appendix~\ref{secA3}, including the worked example in Figure~\ref{fig:appendix_simplicial_complex_example} and Equations~\eqref{homology_k_group}, \eqref{b_2_map}, and \eqref{b_1_map}. The extension of homology theory to encompass the study of these invariant structures across multiple scales is known as \emph{persistent homology} \citep{edelsbrunner2010computational,barcodeGhrist,Carlsson2009TopologyAD}. This extension significantly broadens the scope of homological classification, enabling application to diverse domains such as point clouds, temporal data, and dynamic networks.

\begin{figure}
    \centering
    \includegraphics[width=\linewidth]{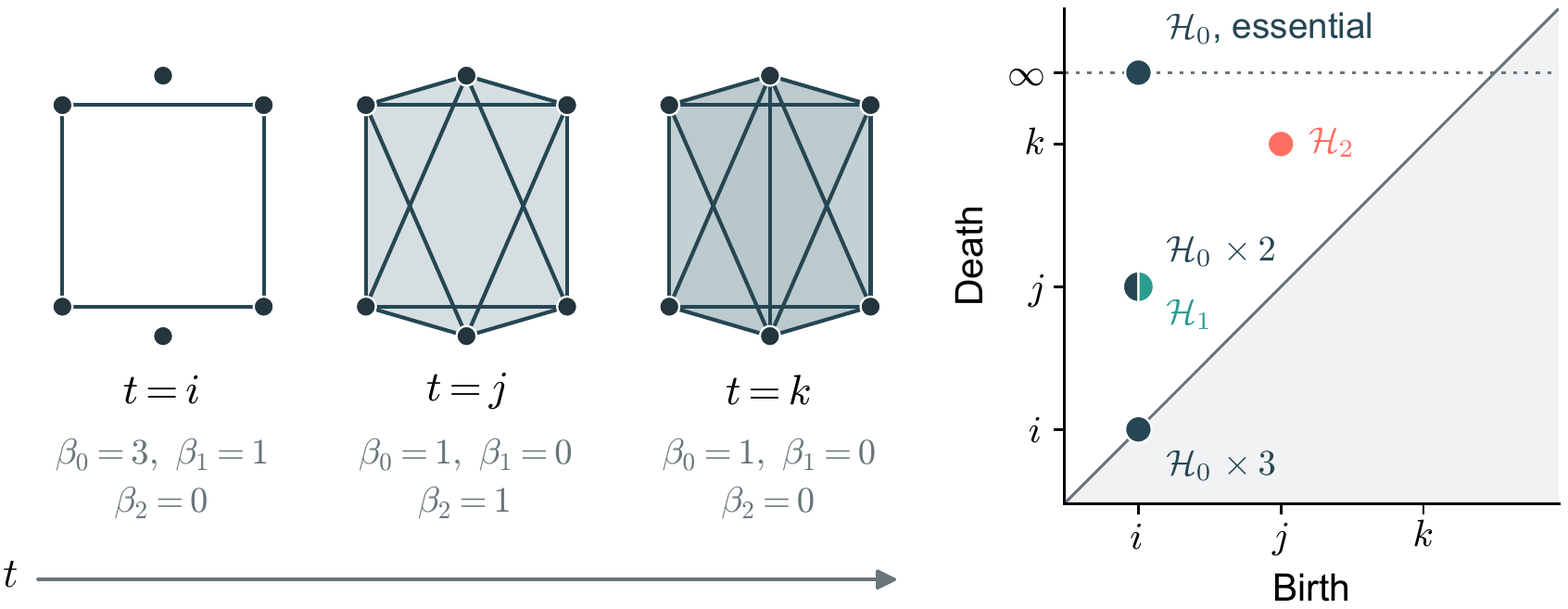}
    \caption{\textbf{A toy example depicting a filtration of a dynamic network (left) and the persistence diagram it produces (right).} The filtration parameter is $t$. At $t=i$ four vertices form a square and two further vertices are isolated; at $t=j$ both of those vertices are joined to all four square vertices; at $t=k$ the two are joined to each other. Every clique (a set of pairwise-connected vertices) is filled, as in the Vietoris--Rips construction used for our network analyses, so at $t=j$ the complex is the boundary of an octahedron. The Betti numbers below each stage are those of the complex drawn above them. In the diagram, each point is labeled with its homological dimension and, where several classes coincide, its multiplicity: three dimension-zero classes are born and lost at $t=i$ as the square closes, two more persist until the isolated vertices join at $t=j$, and one component is essential, meaning that it has no finite death. The square's cycle is born at $t=i$ and is filled at $t=j$; the cavity inside the octahedron is born at $t=j$ and is filled at $t=k$. The split-color marker at $(i,j)$ represents two $\mathcal{H}_0$ classes and one $\mathcal{H}_1$ class at the same birth--death coordinates.}
    \label{fig:persistent_homology_example}
\end{figure}

Consider a network whose nodes and edges are indexed by a scale parameter, such as a distance or correlation cutoff, or publication time as in our application. The parameter orders the network's growth; it need not be temporal. To construct higher-order structures on this network graph, the concept of a \emph{simplicial complex} is introduced, consisting of simple pieces called \emph{simplices}. A $k$-simplex has $k+1$ vertices---a 0-simplex is a vertex, a 1-simplex is an edge, a 2-simplex is a filled-in triangle, and so forth. A \emph{Vietoris--Rips (VR) complex} is a practical method for constructing a simplicial complex from a network. Given nodes of a network, the VR complex adds a $k$-simplex whenever $k+1$ nodes are pairwise within a specified scale parameter $\varepsilon$. For instance, if four nodes are all pairwise within scale $\varepsilon$, then we connect all six edges, fill in each of the four triangles bounded by those edges, and fill in the solid tetrahedron bounded by the four triangles to get a $3$-simplex. The scale at which connections are made can be constructed from any notion of distance between nodes, and results in a nested sequence of VR complexes over this scale parameter. A toy example of persistent homology is shown in Figure \ref{fig:persistent_homology_example}.

Persistent homology is constructed from a \emph{filtration}, a nested sequence of topological spaces (such as the nested VR complexes just mentioned) denoted as $X_1 \subseteq X_2 \subseteq \ldots \subseteq X_N$. The inclusion $X_i \subseteq X_{i'}$ for $i\le i'$ induces a linear map on the $k$th dimensional homology groups $\mathcal{H}_k(X_i) \rightarrow \mathcal{H}_k(X_{i'})$ for all $k\ge 0$. The rank of these homology groups counts the number of distinct $k$-dimensional holes and is called the \emph{$k$th Betti number}, $\beta_k$. Persistent homology traces these topological holes through the filtration, representing them as intervals $[b,d)$ indicating the scale of persistence, where $b$ is the scale at which a hole first appears (is born) and $d$ is the scale at which it no longer remains (dies). Features that never die within the filtration are called \emph{essential} and have death time $d=\infty$. The intervals can be visualized as a \emph{persistence diagram} (PD), a multiset of points in the plane where the $x$-axis indicates the birth coordinate and the $y$-axis is the death. Features are visualized as points $(b,d)$, with points near the diagonal considered short-lived noise, while those further away represent robust topological features.

Comparisons of the topological structure of two filtrations can be made using notions of distance between PDs known as the \emph{$p$-Wasserstein} or \emph{bottleneck distances} \cite{edelsbrunner2010computational}. Stability results specify conditions under which small changes to a network's filtration produce small changes in its persistence diagram \citep{stabilityPD, cohen2010lipschitz, chazal2014persistence}. For example, consider two filtrations of a network with the same nodes, edges, and clique complex, but different entry values. The bottleneck distance between their diagrams in a fixed homological dimension is at most the largest absolute difference between corresponding simplex entry values. Small changes in these entry values therefore produce small changes in the network's persistence diagram.

The bottleneck and Wasserstein distances are computationally intensive metrics and often insufficient for many machine learning (ML) comparison techniques. As such, there has been interest in encoding PDs into more ML-amenable spaces including a functional representation known as a \emph{persistence landscape} \citep{bubenik2015statistical}, a \emph{sliced Wasserstein kernel} \citep{carriere2017sliced}, or a vector in Euclidean space known as a \emph{persistence} image (PI) \citep{adams2017persistence}. We use persistence images in our analyses and provide a brief introduction in the following subsection.

\subsection{Persistence Images}\label{subsec:persistence_images}

At each point in a PD $(b,d)$, we place a probability distribution such as a Gaussian centered at the persistence point (i.e., its mean). Writing $\lambda = d-b$ for the \emph{lifetime} of a persistence point, a common choice weights each point linearly in $\lambda$.

Performing a weighted sum of the distributions over all persistence points, we obtain a \emph{persistence surface} given by
\begin{equation}\label{persistence_image}
    s_B(z) = \sum_{(b,d) \in B} g(b,d) \phi_{(b,d)}(z)
\end{equation}
where $(b,d)$ is a birth-death coordinate in the persistence diagram $B$. We first apply the \textit{birth persistence transformation}, under which the point $(x,y)$ is mapped to $(x,y-x)$, so that each diagram point contributes a Gaussian centered at $(b,\lambda)$ with $\lambda=d-b$, weighted by $\lambda$. The persistence surface is then discretized into a \emph{persistence image} (PI) by evaluating the surface at regularly spaced grid points across the image domain and arranging these values into a two-dimensional array.

Distances between persistence image vectors can be computed several orders of magnitude faster than Wasserstein or bottleneck distances, even accounting for the transformation from a PD to a PI. PIs from multiple dimensions can be concatenated into a single vector. The persistence image construction is also stable with respect to the Wasserstein distance on persistence diagrams, with an explicit Lipschitz bound established in \citep[Theorem 10]{adams2017persistence}. These properties have made PIs a widely used representation of PDs; see for example, \citep{krishnapriyan2020topological, townsend2020representation, NEURIPS2019_12780ea6, bukkuri2021applications, bhaskar2023topological}.

In our analyses, we use a $20\times20$ grid spanning $[0,1]$ in birth and $[0,1.00001]$ in lifetime for every network.  We include essential classes---features with death time $d=\infty$---in the PI by encoding their deaths as \(1.00001\). The finite feature counts and persistence totals reported in the comparison of Section \ref{subsec:thresholding_comparison} exclude these essential classes. We represent each homological dimension with a separate persistence diagram, allowing dimension-specific statistics. We concatenate the vectorized PIs for a given filtration as
\begin{equation}\label{image_decomposition}
    \rho_B = (\rho_{B_1},...,\rho_{B_{k_{\text{max}}}})^T,
\end{equation}
where $k_{\text{max}}$ is the maximal homological dimension, and $\rho_{B_k}$ is the $k$th-dimensional PI corresponding to $B$. We focus on homological features in dimensions $k \geq 1$, including cycles and higher-dimensional cavities. We do not impose a separate constraint on dimension-zero features, which describe connected components.



\subsection{Thresholding Networks of Scientific Concepts}\label{subsec:knowledge_networks}

\begin{figure}[h]
    \centering
    \includegraphics[width=0.66\linewidth]{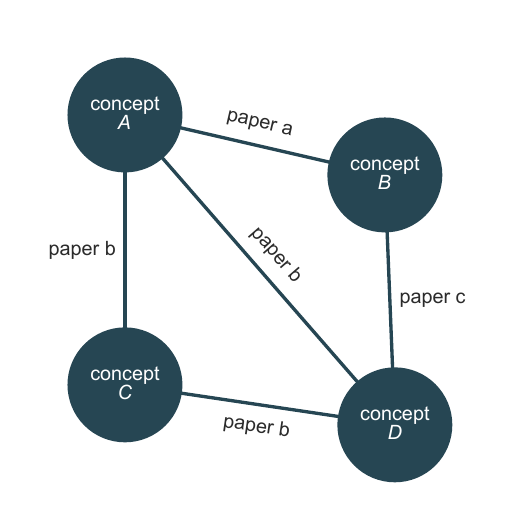}
    \caption{\textbf{A toy concept network on four concepts, labeled $A$, $B$, $C$, and $D$.} These concepts are joined through co-appearance in the abstracts of papers $a$, $b$, and $c$.}
    \label{fig:knowledge_network}
\end{figure}

To concretely illustrate our approach to parameter selection, we apply the pipeline to concept networks in Section \ref{sec:application}. Concept networks are semantic networks where each node corresponds to a scientific concept, and an edge forms between two nodes if the corresponding concepts co-appear in an article abstract. Researchers have used concept networks to study the organization of scientific knowledge within fields \citep{gebhart2020emergence, shi2015weaving, christianson2020architecture}, track how it evolves over time \citep{shi2015weaving, kedrick2024conceptual}, and identify conceptual recombination---the novel pairing of previously unconnected concepts that often precedes discovery and invention \citep{foster2015tradition, fleming2001recombinant, hofstra2020diversity}. We illustrate the basic structure of concept networks using a toy example in Figure \ref{fig:knowledge_network}.

For our application, we are interested in studying how concept networks evolve over time. To capture this temporal dimension, we assign each edge a weight $w\in[0,1]$ based on when two concepts first appear together as
\begin{equation}\label{edge_weight}
    w = \frac{y_\text{publication} - y_{\min}}{y_{\max} - y_{\min}},
\end{equation}
where $y_{\text{publication}}$ is the publication year of the first article in which the two concepts appear together, $y_{\min}$ is the earliest publication year among articles containing at least one concept retained in the network under consideration, and $y_{\max}$ is the latest publication year in the corpus. 


\begin{figure}[h]
    \centering
    \includegraphics[width=0.65 \linewidth]{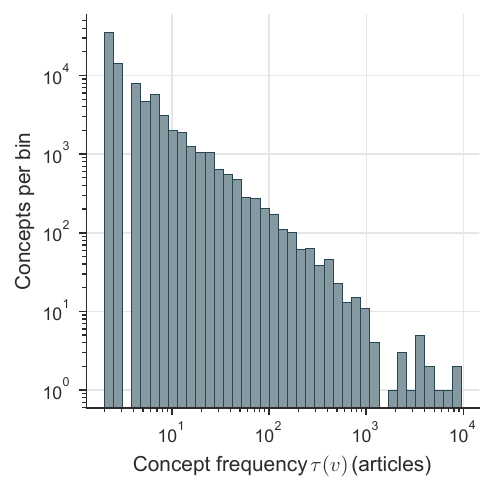}
    \caption{\textbf{Concept frequencies in applied mathematics.} The histogram shows concept frequency $\tau(v)$ using Dimensions AI data for code 0102 of the Australian and New Zealand Standard Research Classification (ANZSRC; see Section \ref{subsec:data_generation} for details). Both axes are logarithmic, with concept frequency on the $x$-axis and the count of concepts per bin on the $y$-axis. While a threshold region can be roughly identified by eye---a lower bound of a few articles, say $\tau(v)=10$, separates the long tail of rare concepts from the rest---small variations in these parameters can produce substantially different networks. A network constructed with lower bound $\ell=10$ may differ considerably from one with $\ell=5$, illustrating the sensitivity of network structure to threshold selection and the need for a principled approach to parameter choice.}
    \label{fig:freqs_dist}
\end{figure}

Concept networks can be thresholded by document frequency to exclude very common and very rare terms. We denote the document frequency of concept $v$ by $\tau(v)$. A typical distribution of $\tau(v)$ is shown in Figure \ref{fig:freqs_dist}. These concepts are typically extracted using natural language processing techniques, often by employing parsers that identify noun phrases (e.g., ``machine learning,'' ``statistical analysis,'' ``climate model''). However, raw network data often include undesirable concepts that introduce noise. Extremely common terms like ``study,'' ``analysis,'' or ``process'' appear frequently but carry little substantive meaning about the conceptual structure of a field. Conversely, very rare terms---including typos (``anlaysis''), highly specialized jargon used in only one or two papers, or artifacts of the extraction process---can fragment the network without contributing meaningful information. Additionally, extraction errors may introduce non-conceptual terms or malformed phrases. We therefore seek to \textit{threshold} the networks using an upper bound $u$ and lower bound $\ell$, which serve as cutoffs on the frequency of concepts such that only those satisfying $\ell \leq \tau(v) \leq u$ are included in the network.

Information-retrieval researchers use term weighting to reduce the influence of common terms in document representations. Term frequency--inverse document frequency (TF-IDF), for example, combines a term's frequency within a document with an inverse-document-frequency factor that decreases as the term appears in more documents. Variants modify term-frequency scaling or normalize document representations \citep{manning2008introduction}. In our concept-network application, we instead use document frequency to decide which concepts to retain as nodes. Document frequency counts the distinct articles containing a concept, so our cutoffs reflect prevalence across articles rather than repeated mentions within an article. We select lower and upper frequency bounds by evaluating the topology of the resulting networks, subject to minimum feature counts.

A common approach to this thresholding problem is to `eyeball' cutoff values based on the frequency distribution. Because network structure may be sensitive to small variations in the parameters (in this case, the upper and lower bounds), studying networks constructed via eyeballing cutoffs can introduce uncertainty in downstream analyses. As an example, in Figure \ref{fig:freqs_dist}, the values of $\ell$ and $u$ that will give the optimal network are not immediately evident.\footnote{Optimal must be defined with respect to some metric, which can vary depending on the goals of the downstream research.} Other thresholding approaches also tend to have the shortcomings discussed in Section \ref{sec:introduction} \citep{curiac, de2010inferring, kawale2013, vandenheuvel}. 

Persistent homology has also been applied to networks by filtering over a correlation or distance threshold \citep{MASUDA20251}. Our filtration parameter is publication year, so the resulting diagram describes when structure appeared rather than how it varies with an edge-weight cutoff. By instead applying the optimization routine defined in Section \ref{sec:algorithm} to the frequency thresholding problem, we can improve on current thresholding methods by accounting for polyadic interactions in the data while still capturing time-dynamic behavior in the network.

\section{Algorithm Details}\label{sec:algorithm}

To improve upon current network thresholding approaches, we develop a method to identify optimal parameterizations based on the stability of the homological structure of a given member of the feature space (here, the space of networks) to variations over the parameter domain. This measure of stability is obtained using the tools introduced in Section \ref{sec:background}. $U$ denotes the parameter space, and $\mathcal{T}:U\to X$ denotes the mapping from parameters to the feature space $X$. We abstract persistent homology as a process $\mathcal{H}:X \to P$ taking an element of $X$ to its persistence diagram(s). The vectorization of a network's persistence diagram(s) via the method of persistence images is denoted as $\rho:P \to \mathbb{R}^n$. Section \ref{mathematical_details_algorithm} defines the local change objective using the image vector at a parameter choice and those at neighboring parameter choices.

\subsection{Algorithm Pipeline}\label{subsec:pipeline_details}
\begin{figure}[h]
    \centering
    \includegraphics[width=0.66\linewidth]{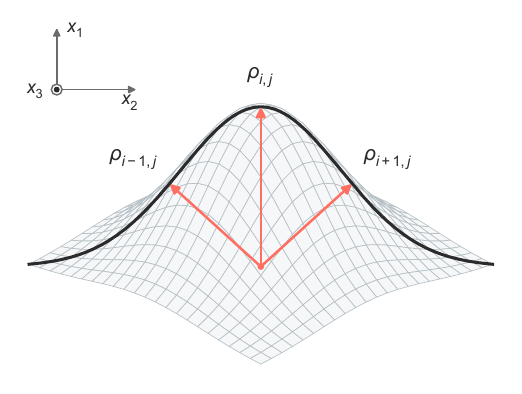}
    \caption{\textbf{Illustration of concatenated PI vectors on a schematic latent surface.} The arrows from the common origin are position vectors to the three representations. The black curve is a one-dimensional trace obtained by varying $i$ while holding $j$ fixed. The magnitudes of the differences $\rho_{i,j} - \rho_{i-1,j}$ and $\rho_{i+1,j} - \rho_{i,j}$ enter the local change calculation.}
    \label{fig:vectorized_manifold}
\end{figure}

We now offer a formal outline of the process, from transforming the data to developing the optimization problem. Details are deferred until Section \ref{mathematical_details_algorithm}. Under $\mathcal{T}$, a specific parameter configuration is mapped to a data representation---for example, requiring a concept to have appeared in at least ten papers may result in a qualitatively different concept network than if the minimum was set to five. Over a subset of $U$---a range of possible parameter or threshold choices---we construct the feature space whose elements are the corresponding transformations of the raw data. Members of this space could be, among others, point clouds or networks, but they should be consistent in construction. Each member $x \in X$ is then assigned to its persistence diagram(s) via $\mathcal{H}$, resulting in a space $P$ of multisets. Each persistence diagram can be transformed into a persistence image vector. Then, for a given network, concatenating the PI vectors from all relevant homological dimensions allows that network to be associated with a single, concatenated PI vector in $\mathbb{R}^n$, where $n$ will depend on the image resolution and the number of topological dimensions being considered. The representations are indexed by the parameter space; for example, an upper and a lower bound define a two-dimensional parameter grid. For such a grid, $i$ and $j$ index its two parameter directions, and $\rho_{i,j}$ denotes the image vector at grid entry $B_{i,j}$. Figure \ref{fig:vectorized_manifold} schematically illustrates a one-parameter trace through these representations.

\begin{figure}
    \centering
    \includegraphics[width=\linewidth]{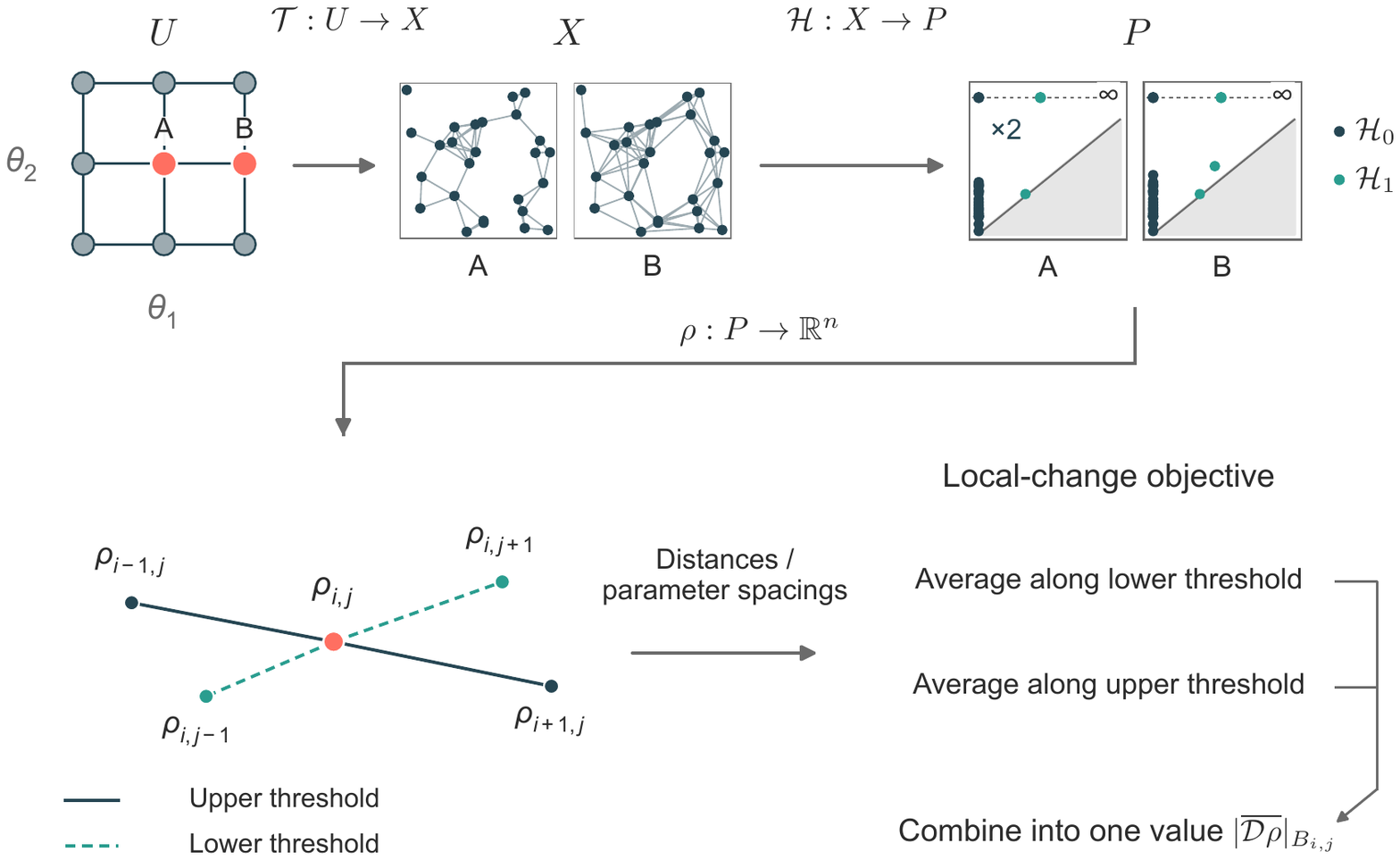}
    \caption{ \textbf{Visualization of the algorithm's pipeline, excluding the optimization step.} Parameter choices in the evaluated grid within $U$ determine networks in the feature space $X$, which persistent homology $\mathcal{H}$ transforms into persistence diagrams in $P$. Each network has an associated representation in $\mathbb{R}^n$ by the process of persistence images, $\rho$. We compare these representations at neighboring parameter choices to evaluate the local change objective. Matching A/B labels identify the illustrated parameter settings, toy networks, and persistence diagrams. In the PDs, birth is horizontal and death is vertical; charcoal and verdigris denote $\mathcal{H}_0$ and $\mathcal{H}_1$, respectively. Dashed lines marked $\infty$ show classes with no finite death, and $\times2$ records multiplicity. The bottom row illustrates an interior parameter-space grid point and its four neighbors. The normalized image distances in Equation \eqref{discrete_grad} are averaged along each parameter direction and combined into the objective value $|\overline{\mathcal{D}\rho}|_{B_{i,j}}$ in Equation \eqref{average_directional_derivative}, expressed for the two threshold indices in Equation \eqref{objective_function_application}. The displayed image-vector positions are schematic.}
    \label{fig:pipeline}
\end{figure}

We measure local change by comparing the PI vector at a parameter choice with the PI vectors at its immediate parameter-grid neighbors. We divide each distance by the corresponding parameter separation, average these normalized distances separately along each parameter direction, and combine the averages into the objective value for that parameter choice. Smaller objective values indicate less change in the topological representation under the neighboring threshold choices. This pipeline is shown in Figure \ref{fig:pipeline}.

\subsection{Algorithm Properties}\label{mathematical_details_algorithm}
We measure local change by comparing persistence image vectors at neighboring entries of an evaluated parameter grid. Labels such as $A$, $B$ and $C$ identify grid entries, and $\rho_B$ denotes the concatenated PI vector at entry $B$. Distinct entries may yield the same network or image vector. Two entries are immediate neighbors in a parameter direction when all other coordinates agree and their values in that direction are consecutive in its ordered grid values. We include such a neighbor only when that grid entry has been evaluated. The coordinate grids need not be equally spaced. We use $p=2$ in all norms in the local change calculation. For immediate neighbors $B$ and $B'$ in a parameter direction $\theta$, let $\theta_B$ and $\theta_{B'}$ denote their values in that direction. We define the normalized persistence image distance between them by
\begin{equation}\label{discrete_grad}
    |\mathcal{D} \rho|_{B,B'} = \frac{\hphantom{|}||\rho_B - \rho_{B'}||_p}{|\theta_B - \theta_{B'}|} \ .
\end{equation}

By taking the average of Equation \eqref{discrete_grad} over the available immediate neighbors separately in each parameter direction, and then taking the Euclidean norm of these coordinate averages, we obtain the local change objective $|\overline{\mathcal{D} \rho}|_B$. For example, if the parameter space can vary over a bounded subset of $\mathbb{R}^2$, then after discretizing this subset into a complete grid of desired refinement, an interior entry $B$ will have four neighbors; two will be in the $\theta_1$-direction, which we denote as $A$ and $C$, and two will be in the $\theta_2$-direction, which we denote as $D$ and $E$. The stability of $B$ would then be measured according to
\begin{samepage}
\begin{gather}
    |\overline{\mathcal{D} \rho}|_B = \bigg|\bigg| \frac{\hphantom{2,}||\rho_B-\rho_A||_p}{2|\theta_{1,B} - \theta_{1,A}|}\hat{\textbf{e}}_{\theta_1} + \frac{\hphantom{2,}||\rho_B - \rho_C||_p}{2|\theta_{1,B}-\theta_{1,C}|}\hat{\textbf{e}}_{\theta_1} + \nonumber \\\frac{\hphantom{2,}||\rho_B - \rho_D||_p}{2|\theta_{2,B}-\theta_{2,D}|}\hat{\textbf{e}}_{\theta_2} + \frac{\hphantom{2,}||\rho_B - \rho_E||_p}{2|\theta_{2,B}-\theta_{2,E}|}\hat{\textbf{e}}_{\theta_2} \bigg|\bigg|_p \label{average_directional_derivative}
\end{gather}
\end{samepage}

where $\hat{\textbf{e}}_{\theta_i}$ is the unit vector in the $\theta_i$-direction. The value of two in the denominator ensures that interior values can be compared against boundary values; if $B$ instead occurs on a border of the grid, we can no longer average in the direction in which the border occurs and therefore give full weight to the only occurring neighbor in that direction. For example, if $B$ had two neighbors $A$ and $C$ in the $\theta_1$-direction but only one neighbor $D$ in the $\theta_2$-direction, Equation \eqref{average_directional_derivative} is recast as
\begin{equation}\label{edge_case_average_directional_derivative}
    |\overline{\mathcal{D} \rho}|_B = \bigg|\bigg| \frac{\hphantom{2,}||\rho_B-\rho_A||_p}{2|\theta_{1,B} - \theta_{1,A}|}\hat{\textbf{e}}_{\theta_1} + \frac{\hphantom{2,}||\rho_B - \rho_C||_p}{2|\theta_{1,B}-\theta_{1,C}|}\hat{\textbf{e}}_{\theta_1} + \frac{\hphantom{2,}||\rho_B - \rho_D||_p}{|\theta_{2,B}-\theta_{2,D}|}\hat{\textbf{e}}_{\theta_2} \bigg|\bigg|_p \, .
\end{equation}

We optimize according to the constrained minimization problem  
\begin{align}\label{optimization_step}
\begin{split}
    \argmin{i \in I} \ & |\overline{\mathcal{D} \rho}|_{B_i}\\ \text{subject to }&  f^k_i \geq \delta_k , \text{ for all }\, k=1,...,k_{\text{max}},
\end{split}
\end{align}
where $I$ indexes the parameter-grid entries $B_i$ eligible for selection, $f_i^k$ is the number of $k$-dimensional homological features at grid entry $B_i$, and $|\overline{\mathcal{D} \rho}|_{B_i}$ is the local change objective value for that grid entry. Evaluated entries outside $I$ may still supply neighboring image vectors. The hyperparameters of the algorithm are the $k_{\text{max}}$ constraints $\delta_k$, where $k_{\text{max}}$ is the maximal homological dimension computed by $\mathcal{H}$. These constraints can be viewed as user-prescribed requirements for the number of $k$-dimensional features that must be present in the optimal representation.

\begin{samepage}
We implement our topological thresholding algorithm in HONS (higher-order network selection), a Python package available at \url{https://github.com/rfunklab/hons}. Our companion replication repository for this study (\url{https://github.com/rfunklab/hons-replication}) provides persistence diagrams and network-level measurements computed for the analyses in Section~\ref{sec:application}, together with scripts for reproducing the threshold selections, higher-order variance, and method-comparison statistics.

\end{samepage}

\section{Application}\label{sec:application}
In this section, we apply topological thresholding to concept networks constructed from scientific abstracts. We use concept networks from applied mathematics to illustrate the construction of the threshold search space, examine the selected networks, and assess how the network selected for local stability compares with topological variation across the threshold grid. We then compare topological thresholding with ten alternative thresholding methods, first in applied mathematics and then across scientific fields.

\subsection{Data}\label{subsec:data_generation}
We build the feature space of concept networks using data from Dimensions AI \cite{herzog2020dimensions}, a comprehensive index of over 130 million publications across all fields of science. Dimensions AI uses machine learning and natural language processing to extract structured metadata from scholarly documents, including publication metadata, citation links, and concept annotations. We use a snapshot of the database from September 1, 2021, focusing on publications from 1920 to 2021 to capture a century of conceptual evolution (the final year is partial, given the snapshot date).

We develop the detailed application using 167,001 articles in applied mathematics, classified by Australian and New Zealand Standard Research Classification (ANZSRC) code 0102. Appendix~\ref{SecA4} presents an additional application in zoology (Figures~\ref{fig:zoology_manifold}, \ref{fig:zoology_results}, and \ref{fig:zoology_variance}). Section~\ref{subsec:thresholding_comparison} evaluates comparative performance across scientific fields; Appendix~\ref{app:comparison_methods} specifies the input preparation for the detailed example and the cross-field evaluation.

Dimensions AI's concept extraction employs natural language processing techniques to identify key scientific concepts from article abstracts. Each concept in our dataset includes several attributes, including the associated article identifier, the year of publication, the concept itself (typically a noun phrase such as ``differential equation'' or ``numerical method''), a relevance score (a measure of the concept's pertinence to the given article), and the frequency $\tau(v)$ with which the concept appears across the entire corpus. From this data, we construct concept networks as described in Section \ref{subsec:knowledge_networks}, where nodes represent concepts, and edges connect concepts that co-appear in the same article abstract. Before any thresholding, the applied mathematics network contains $81{,}026$ concepts and $1{,}361{,}809$ co-occurrence edges.

\begin{figure}
    \centering
    \includegraphics[width=0.66\linewidth]{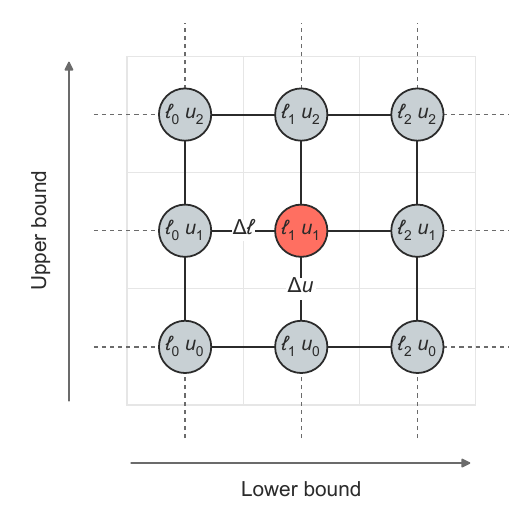}
    \caption{\textbf{A threshold choice and its immediate neighbors in parameter space.} Each marker denotes a threshold pair and corresponds to the network constructed with those bounds. The focal network is constructed at parameter values $\ell_1$ and $u_1$. The parameter `directions' are indicated by $\Delta \ell$ and $\Delta u$, and the direction of positive change (increase in cutoff value) is indicated by the arrows along the sides.}
    \label{fig:network_space}
\end{figure}

Network thresholding is performed on parameters defining bounds on concept document frequency. The parameter space $U$ therefore consists of pairs of parameters $(\ell,u)$, where $\ell$ is a lower bound on concept document frequency, and $u$ is an upper bound on concept document frequency. That is, for a given set of parameters, $(\ell, u)$, a concept is only included in the concept networks if it appears in at least $\ell$ papers and at most $u$ papers. The feature space $X$ is the collection of the networks generated by all unique combinations of $\ell$ and $u$, whose values are given in Appendix \ref{secA2}. The $\ell$ and $u$ values are selected from a $9\times 16$ Cartesian product, resulting in 144 distinct grid entries i.e. networks. The column corresponding to the smallest lower bound is excluded from the optimization, resulting in 128 selectable networks. A parameter-grid neighborhood and its corresponding networks are illustrated in Figure \ref{fig:network_space}. We compare the network at each parameter choice with those at its immediate parameter-grid neighbors; for example, in Figure \ref{fig:network_space}, the neighbors of the network corresponding to the parameter pair $(\ell_1,u_1)$ are the networks corresponding to the parameter pairs $(\ell_0,u_1)$, $(\ell_1,u_0)$, $(\ell_1,u_2)$, and $(\ell_2,u_1)$. 

After assigning the edge weights in each network according to Equation \eqref{edge_weight}, we implement persistent homology on each $x \in X$ using a filtration parameter $ \varepsilon \in [0,1]$, adding edges to the network whenever $w \leq \varepsilon$. In this application, we use homological features in dimensions one and two, corresponding to cycles and cavities. The constraints $\delta_1$ and $\delta_2$ specify the minimum numbers of features required in those dimensions. We impose no separate constraint on connected components. The persistence diagrams reported here are computed with the Python package Geometry Understanding in Higher Dimensions (GUDHI) \cite{gudhi:urm}. We evaluate the objective at every grid entry satisfying the feature constraints and select an entry with the smallest value. Ties are recorded when they occur. 

\begin{figure}[h]
    \centering

    \includegraphics[width=\linewidth]{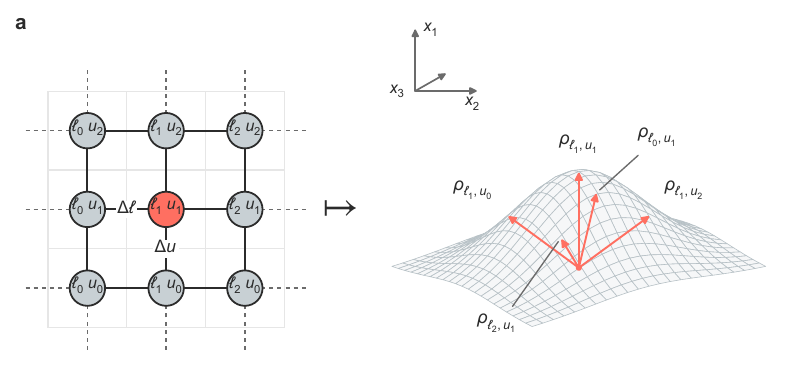}\\[6pt]
    \includegraphics[width=0.8\textwidth]{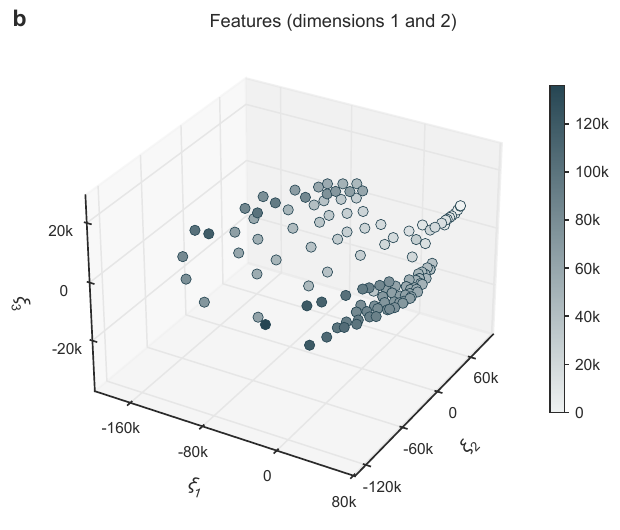}
    
    \caption{\textbf{Persistence-image representations of candidate networks.} (a) Illustration of how the networks associated with neighboring threshold choices in Figure \ref{fig:network_space} map to PI representations. The arrows from the common origin are position vectors to those representations; changes in $\ell$ or $u$ select neighboring grid points. (b) Each point represents an applied mathematics network's concatenated PI vector, projected onto the first three principal components ($\xi_1$, $\xi_2$, $\xi_3$) and colored according to the sum of the dimension one and two feature counts. Figures \ref{fig:fig_2_reworked} and \ref{fig:varied_tuning_parameters} locate the selected networks on the underlying threshold grid.}
    \label{fig:networks_correlation_to_surface}
\end{figure}

We compute a concatenated persistence image vector for each evaluated threshold pair. Because $U$ varies in two directions, we combine normalized differences along both parameter directions. For an interior grid entry, the objective defined in Equation \eqref{average_directional_derivative} and minimized in Equation \eqref{optimization_step} is
\begin{gather}
    |\overline{\mathcal{D} \rho}|_{B_{i,j}} = \bigg|\bigg|\frac{\hphantom{2,}||\rho_{i,j}-\rho_{i+1,j}||_p}{2|u_{i,j} - u_{i+1,j}|}\hat{\textbf{e}}_u + \frac{\hphantom{2,}||\rho_{i,j} - \rho_{i-1,j}||_p}{2|u_{i,j}-u_{i-1,j}|}\hat{\textbf{e}}_u + \nonumber\\\frac{\hphantom{2,}||\rho_{i,j} - \rho_{i,j+1}||_p}{2|\ell_{i,j}-\ell_{i,j+1}|}\hat{\textbf{e}}_\ell + \frac{\hphantom{2,}||\rho_{i,j} - \rho_{i,j-1}||_p}{2|\ell_{i,j}-\ell_{i,j-1}|}\hat{\textbf{e}}_\ell \bigg|\bigg|_p
\label{objective_function_application}
\end{gather}
where the indices $i$ and $j$ correspond to the parameter pair $(\ell_j,u_i)$, with $\ell_{i,j}=\ell_j$ and $u_{i,j}=u_i$, and $\rho_{i,j}$ is the concatenated PI vector at that entry. When only one immediate neighbor is available in a parameter direction, we use the boundary rule in Equation \eqref{edge_case_average_directional_derivative}. If no upper-bound neighbor is available, that direction contributes zero. Every eligible entry in this application has a lower-bound neighbor. The lowest lower-bound column is evaluated to supply neighboring image vectors but is excluded from selection in Equation \eqref{optimization_step}. Figure \ref{fig:networks_correlation_to_surface}a illustrates the correspondence between neighboring threshold pairs and their image vectors; labels of the form $\rho_{\ell,u}$ identify vectors by their threshold values.

We compute the persistence images by evaluating the weighted Gaussian surface at fixed grid points. In Equation \eqref{persistence_image}, each feature is weighted by its lifetime $\lambda=d-b$, so features with greater persistence receive greater weight. For $\phi$, we use a rotationally symmetric Gaussian distribution with standard deviation $\sigma=0.1$. The rotational symmetry ensures that the Gaussian does not preferentially weight either the birth or persistence direction. We evaluate the surface at the $20\times20$ evenly spaced points over $[0,1]\times[0,1.00001]$ described in Section~\ref{subsec:persistence_images}. For each threshold pair $(\ell,u)$, we concatenate the dimension one and dimension two PI vectors, each with $400$ entries, to obtain a vector in $\mathbb{R}^{n}$ with $n=800$.

Our choices of Gaussian standard deviation and PI resolution are guided by the empirical parameter sensitivity analysis in Adams et al. \cite{adams2017persistence}, which found relatively little variation in classification accuracy across a broad range of hyperparameters. We also assess these two choices directly on the applied mathematics grid. Holding the persistence diagrams and every other element of the construction fixed, we recompute the selections over the Cartesian product of $\sigma\in\{0.05,0.1,0.2\}$ and resolutions $10\times10$, $20\times20$ and $40\times40$. We keep the feature count constraints fixed across configurations. Because the PI settings do not change the feature counts, the same networks satisfy those constraints in every configuration. For each of the three illustrated constraint settings of Figure \ref{fig:fig_2_reworked}, the selected network is unchanged across all nine PI configurations. Selections are also unchanged across the three resolutions tested at each fixed value of $\sigma$. Table~\ref{tab:image_sensitivity} in Appendix~\ref{secA2} reports the comparison in full. 

For illustration, we project the applied mathematics PI vectors onto the first three principal components in Figure \ref{fig:networks_correlation_to_surface}b. Positions summarize differences between those vectors, while color shows the sum of the dimension one and two feature counts.


\subsection{Results}\label{subsec:results}

\begin{figure}
    \centering
    \includegraphics[width=1\linewidth]{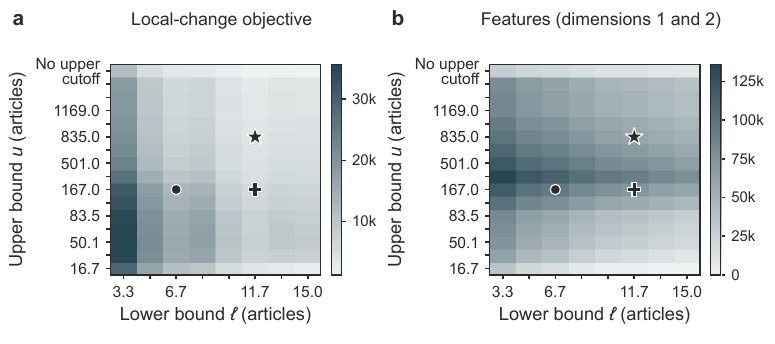}
    \caption{\textbf{Optimal networks under three feature-count constraint settings.} Panels (a) and (b) show the local change objective and the sum of the dimension one and two feature counts, respectively, across the applied mathematics threshold grid. The objective combines the normalized image distances averaged in both the $\ell$ and $u$ directions. The color bars are on different scales. The black star, plus sign, and dot mark the selections for $(\delta_1,\delta_2)=(P_1(25),P_2(25))$, $(P_1(50),P_2(25))$, and $(P_1(75),P_2(50))$, respectively. Here, $P_k(q)$ is the $q$th percentile of the number of $k$-dimensional features taken over the 128 selectable networks of the grid, generated on the applied mathematics data from Dimensions AI.}
    \label{fig:fig_2_reworked}
\end{figure}

To demonstrate the algorithm's performance, we display results for three different combinations of the $\delta_1$ and $\delta_2$ constraints in Figure \ref{fig:fig_2_reworked}. We set $\delta_k$ to a percentile of the distribution of $k$-dimensional feature counts over the 128 selectable networks of the grid, writing $\delta_k = P_k(q)$ for the $q$th percentile. The three configurations considered are $(\delta_1, \delta_2) = (P_1(25), P_2(25))$, $(P_1(50), P_2(25))$, and $(P_1(75), P_2(50))$. The left column shows the local change objective. The image in the right column visualizes the distribution of the sum of the dimension one and two homological features as $\ell$ and $u$ vary. The optimal selections are identified by markers in the plots; we note that the optimum lands in regions of relatively lower values of $|\overline{\mathcal{D} \rho}|$ for all three cases. Furthermore, increasing the constraints shifts the selection toward networks with more homological features. Increasing the constraints from the $(P_1(25), P_2(25))$ to the $(P_1(75), P_2(50))$ configuration results in a much smaller lower bound $\ell$; a smaller lower bound admits concepts appearing in fewer articles, while it is the upper bound $u$ that governs whether the most frequent concepts are retained.

To better understand the effects of the tuning hyperparameters, we run the optimization algorithm for a series of 100 unique $\delta_1$ and $\delta_2$ combinations, where we generate the series according to the Cartesian product of ten increasingly restrictive $\delta_1$ and ten increasingly restrictive $\delta_2$ choices, listed in Table~\ref{tab:hyperparameters} of Appendix~\ref{secA2}. These results are shown in Figure \ref{fig:varied_tuning_parameters}. Ninety-nine of the 100 settings admit a feasible network; the most restrictive combination, $(P_1(91),P_2(91))$, admits none. We measure a setting's restrictiveness by the sum of its two constraint percentiles. Each bubble marks a selected cell (corresponding to a network), its area giving the number of parameter settings that select it, and its shade the mean restrictiveness of those settings, so repeated selections appear as larger bubbles. The 99 feasible settings concentrate on 19 distinct cells. In general, increasing $\delta_1$ and $\delta_2$ moves the selection toward cells with more homological features. The nine strictest feasible settings select four distinct cells. The high-objective ridge is the leftmost column displayed in the left plot, at $\ell=3.3$, whose mean objective is about 1.8 times that of the next column; the algorithm mostly avoids it, with four of the 99 feasible settings selecting a cell there.

The feature-count constraints let users specify minimum numbers of cycles and voids while the objective selects for local stability. In our study, we search the entire selectable grid and choose the network with the smallest local change objective among candidates satisfying both feature-count requirements, as specified in Equation \eqref{optimization_step}. In applied mathematics, the three illustrated constraint settings of Figure \ref{fig:fig_2_reworked} admit 71, 44, and 10 of the 128 selectable networks. These counts illustrate how stricter requirements narrow the set of networks available for the stability comparison.

For a new application, we suggest that users begin with $\delta_1=P_1(50)$ and $\delta_2=P_2(25)$, the setting used in our method comparisons. These requirements specify that the selected network must contain at least the median number of cycles and the 25th-percentile number of voids across the selectable candidate networks. Users should examine how neighboring percentile settings change the selected thresholds and the numbers of cycles and voids in the resulting networks, assessing whether those networks suit their intended analysis. Users should report the adopted percentiles and resulting feasible-set size.

\begin{figure}
    \centering
    \includegraphics[width=1\linewidth]{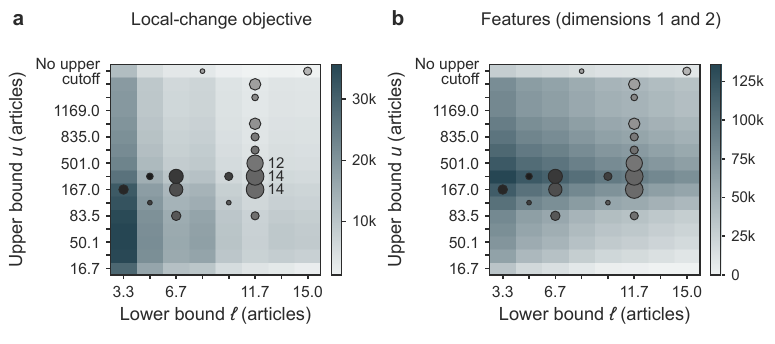}
    \caption{\textbf{The 99 feasible constraint settings select 19 distinct grid cells.} We evaluate 100 percentile combinations of $\delta_1$ and $\delta_2$ on the same applied mathematics threshold grid as Figure \ref{fig:fig_2_reworked}. Background shading shows (a) the local change objective and (b) the sum of the dimension one and two feature counts; the color bars describe these background quantities. Each bubble marks a selected grid cell; its area is the number of constraint settings that select that cell and its shade is the mean sum of the two constraint percentiles across those settings, with lighter tones for looser constraints. Numbers label the selection counts for the three most frequently selected cells. Of the 100 settings tested, 99 admit a feasible cell and one, $(P_1(91),P_2(91))$, admits none, because no candidate network reaches both percentile requirements.}
    \label{fig:varied_tuning_parameters}
\end{figure}

\begin{figure}[h]
    \centering
    \includegraphics[width=\linewidth]{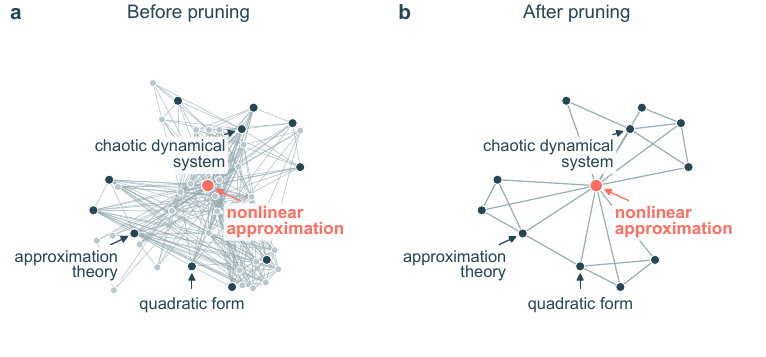}
    \caption{\textbf{A concept-network neighborhood before and after thresholding.} Subfigure (a) shows the neighborhood induced on the concept `nonlinear approximation' in the unthresholded network: 41 concepts joined by 281 co-occurrence links. Subfigure (b) shows the same neighborhood after applying the bounds of the primary selection in Figure \ref{fig:fig_2_reworked} (black plus sign), leaving 12 concepts and 23 links. Those bounds, selected on the full applied mathematics network, retain concepts appearing in 12--167 articles. Surviving concepts occupy the same positions in both panels. Coral marks the focal concept, charcoal marks the other retained concepts, and light gray marks concepts removed by thresholding. The concepts removed include general terms above the upper bound, such as `differential equations' and `optimization problem'; those retained are more specific, such as `chaotic dynamical system' and `convex optimization'. The neighborhood was chosen for legibility.}
    \label{fig:comparison_focal_networks}
\end{figure}

In Figure \ref{fig:comparison_focal_networks}, we compare a subgraph in the original applied mathematics network to the same subgraph after applying the cutoffs $\ell=11.69007$ papers and $u=167.001$ papers, corresponding to the optimal selection for $(\delta_1,\delta_2) = (P_1(50), P_2(25))$. Because document frequencies are integer counts, these cutoffs retain concepts appearing in 12--167 articles, inclusive. We apply the pruning algorithm and threshold selection to the full applied mathematics network; the subgraph shown here is selected solely as an illustrative example, since the full network is too large to visualize effectively. The neighborhood shown is centered on `nonlinear approximation', a concept appearing in 15 articles, and was chosen for legibility in both panels.  The neighborhood falls from 41 concepts and 281 co-occurrence links to 12 concepts and 23 links. The concepts removed here are the corpus-wide general terms that exceed the upper bound, such as `differential equations', `partial differential equation' and `optimization problem', together with concepts below the lower bound; those retained are more specific, such as `chaotic dynamical system', `convex optimization' and `approximation theory'. Retaining only nodes within the selected frequency band leaves a sparser structure that is more amenable to further analysis such as feature extraction.

\subsection{Relating Local Stability to Topological Variability across Candidate Networks}\label{sec:statistical_implications}

In this section, we examine how the network selected for local stability compares with the broader topological variation across the threshold grid. Stability under nearby threshold changes does not tell us how the selected network compares with candidates elsewhere in that space. We therefore ask whether minimizing local change also yields a network with low topological deviation when assessed against the candidate grid as a whole.

To evaluate how each network's relative representation of cycles and voids compares with the grid as a whole, we define a measure of higher-order variance. Because thresholding can affect one-dimensional cycles and two-dimensional voids differently, we focus on how their relative representation varies across candidate networks. For each candidate network, we take the difference between its one- and two-dimensional persistence images at corresponding birth and lifetime coordinates. This difference captures the contrast between the two kinds of features, accounting for when they appear and how long they persist. Averaging these differences across the grid provides a common central reference for a squared-deviation comparison. Higher-order variance is the scaled squared distance of each network's image difference from this reference. Lower values indicate a relative pattern of cycles and voids closer to the grid average. Among networks satisfying the same feature-count constraints, we compare the locally stable selection with those having the smallest departures from this reference.

For a network with persistence diagram $B$, let $\rho_{B_k}$ denote its $k$th-dimensional persistence image vector, as in Equation \eqref{image_decomposition}, and let $\overline{\rho_k}$ denote the componentwise mean of that image across all networks in the grid. For the two homological dimensions considered here, its higher-order variance is
\begin{align}\label{h_o_var}
\begin{split}
    V_2&(B) :=\\ &\frac{1}{\eta-1} \sum_{i=1}^{\eta} \bigg[ 
    (\rho_{B_1}^{(i)} - \overline{\rho_1}^{(i)})^2 + (\rho_{B_2}^{(i)} - \overline{\rho_2}^{(i)})^2 - 2(\rho_{B_1}^{(i)} - \overline{\rho_1}^{(i)})(\rho_{B_2}^{(i)} - \overline{\rho_2}^{(i)})
    \bigg]
\end{split}
\end{align}
where $n/2=\eta=400$ is the number of pixels in a single-dimension persistence image and the superscript $i$ indexes those pixels; $\overline{\rho_k}$ is the componentwise mean of the $k$th-dimensional image taken across all $m=144$ networks of the grid. We describe motivations for Equation \eqref{h_o_var}, as well as an alternative variance measure that generalizes to $k_{\text{max}}$ dimensions, in Appendix \ref{sec:notes_on_variance}.

\begin{figure}
    \centering
    \includegraphics[width=\linewidth]{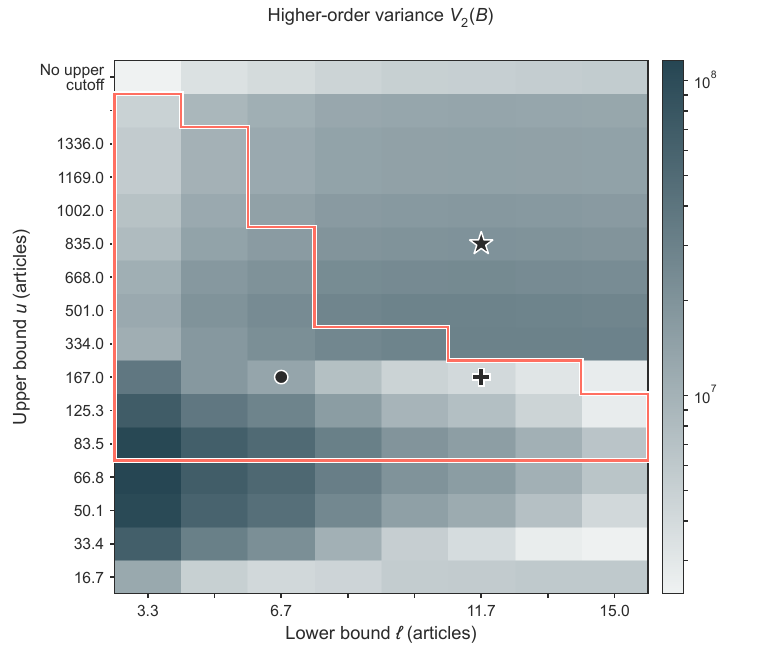}
    \caption{\textbf{Higher-order variance over the applied mathematics network space.} Values are evaluated for $k_{\text{max}}=2$ using Equation \eqref{h_o_var} and the networks described in Section \ref{subsec:results}. The coral outline encloses the 44 networks satisfying the primary constraints $(\delta_1,\delta_2)=(P_1(50),P_2(25))$. The primary selection (black plus sign) has the third-lowest value among these 44 networks and the tenth-lowest among the 128 selectable networks. The selections at $(P_1(25),P_2(25))$ (star) and $(P_1(75),P_2(50))$ (dot) rank ninety-first and fifty-seventh among the 128 selectable networks.}
    \label{fig:variance}
\end{figure}

We compute higher-order variance for each concept network from Section \ref{subsec:results} (Figure \ref{fig:variance}). The primary selection, at $(\delta_1,\delta_2)=(P_1(50),P_2(25))$, has the third-lowest value among the 44 networks satisfying these constraints and the tenth-lowest among the 128 selectable networks. The two feasible networks with lower variance have larger local change objectives. The primary selection therefore minimizes local topological sensitivity over the feasible grid while also having low higher-order variance, despite the latter not being an optimization objective.

\clearpage
\subsection{Comparison with Other Thresholding Methods}\label{subsec:thresholding_comparison}

In this section, we assess whether and to what degree topological thresholding outperforms alternative network thresholding methods in retaining higher-order structure. We first examine feature densities and lifetime distributions in the applied mathematics network introduced above, then evaluate the same aggregate measures across dozens of scientific fields. For each input network, all methods begin with the same nodes and edges, but we measure the number and persistence of homological features in the networks that each method retains.

The comparison includes simple cutoffs on node properties and edge weights, together with statistical methods that account for differences among nodes. Fixed node-based thresholds select nodes according to a scalar property; in the concept network, that property is document frequency, the number of articles containing a concept. We test minimum frequencies of two and five articles and the frequency interval of 10--150,000 articles. Global edge-weight thresholding instead applies one cutoff to all edges~\cite{serrano2009extracting,foti2011}.  Here, the selection weight is the number of articles containing both endpoint concepts, and we retain edges with counts of at least two or at least five.

Statistical backbone methods assess edge weights relative to local distributions or expectations based on node frequencies or strengths (sums of incident edge weights). These methods can retain an edge whose weight is small in absolute terms but large relative to the weights expected at its endpoints. We evaluate five published thresholding methods---disparity~\cite{serrano2009extracting}, locally adaptive network sparsification (LANS)~\cite{foti2011}, marginal likelihood thresholding~\cite{dianati2016}, noise-corrected thresholding~\cite{coscia2017}, and hypergeometric thresholding~\cite{tumminello2011}. We use $\alpha=0.05$ for disparity, LANS, marginal likelihood thresholding, and hypergeometric thresholding, and the corresponding normal-quantile cutoff for the variance-based noise-corrected threshold. Appendix~\ref{app:comparison_methods} specifies the thresholding criteria and retention rules.

We hold the topological thresholding constraints fixed at $(\delta_1,\delta_2)=(P_1(50),P_2(25))$ throughout the comparison, using the configuration introduced in Section~\ref{subsec:results}. These constraints require the one- and two-dimensional feature counts to reach at least the 50th and 25th percentiles, respectively, of their distributions across candidate networks. Among the qualifying networks, the method selects the optimum thresholds with the smallest local change in the persistence image representation. In applied mathematics, the selected bounds are $\ell=11.69007$ and $u=167.001$, retaining concepts that appear in 12--167 articles.

We evaluate all retained networks using the temporal filtration employed in the preceding analysis, which follows the formation and filling of homological features over publication time. We convert the normalized years (as described in Equation~\ref{edge_weight}) to calendar years to enable direct comparisons of lifetimes. Thresholding uses document frequencies and co-occurrence counts from the full corpus, whereas a retained edge enters the filtration in the (normalized) year when its endpoint concepts first co-occur. Thus, an edge selected by a count threshold enters at its first occurrence, even if that count is reached in a later year. Triangles and tetrahedra enter when their last constituent edge appears, yielding the temporal clique complex used to compute $\mathcal{H}_1$ and $\mathcal{H}_2$. For the comparison measures, we count positive-lifetime features whose formation and filling both occur within the observation period, so every counted feature has a measured lifetime in calendar years. Features that remain open at the end of the filtration are excluded from these measures.

We report feature densities to compare higher-order structure relative to the elements from which that structure is formed. One-dimensional homological features are represented by cycles of edges, while two-dimensional features are represented by closed combinations of triangles. We therefore divide the finite $\mathcal{H}_1$ count by the number of edges and the finite $\mathcal{H}_2$ count by the number of triangles in the final retained graph, multiplying each ratio by 1,000. To account for how long the features persist, we also sum their lifetimes and normalize those sums by the same edge or triangle counts. The resulting normalized total finite persistence has units of years per 1,000 edges for $\mathcal{H}_1$ and years per 1,000 triangles for $\mathcal{H}_2$. A larger value can result from more features, longer lifetimes, or both.

\paragraph{Detailed Comparison}
Our topological thresholding method yields higher finite feature densities and normalized total finite persistence than all ten alternatives in the applied mathematics network (Figure~\ref{fig:thresholding_comparison}a--d). The retained network contains 84.3 finite $\mathcal{H}_1$ features per 1,000 edges, compared with 21.9 for marginal likelihood, the highest alternative value. For $\mathcal{H}_2$, topological thresholding yields 15.1 finite features per 1,000 triangles, compared with 5.75 for the noise-corrected threshold. These correspond to density ratios of 3.85 and 2.63. Relative to the same alternatives, normalized total finite persistence is 4.20 times as large for $\mathcal{H}_1$ and 1.45 times as large for $\mathcal{H}_2$. Table~\ref{tab:comparison_applied_math} in Appendix~\ref{app:comparison_results} reports network sizes, finite feature counts, and all four topological measures. Because the corpus retains only concepts appearing in at least two articles, the document-frequency-$\geq2$ threshold leaves this network unchanged and coincides with the unthresholded input.

The lifetime profiles show that the higher feature densities do not include just short-lived features (Figure~\ref{fig:thresholding_comparison}e,f). Each curve gives the density of finite features lasting at least the number of years on the horizontal axis. Topological thresholding has the highest density for minimum lifetimes through 49 years in $\mathcal{H}_1$ and 6 years in $\mathcal{H}_2$; beyond those cutoffs other methods yield higher densities. The two dimensions reach their aggregate advantage differently. In $\mathcal{H}_1$, topological thresholding combines the higher feature density with a longer mean finite lifetime than marginal likelihood, 10.05 against 9.20 years. In $\mathcal{H}_2$, its mean finite lifetime is shorter than that of noise-corrected thresholding, 3.31 against 6.02 years, so the aggregate advantage in this case rests on feature density rather than on individual features lasting longer.

\clearpage
\newgeometry{margin=0.5in,includefoot}
\newsavebox{\figTwelveBox}
\newcommand{\figTwelveCaption}{\textbf{In applied mathematics, topological thresholding exceeds ten alternatives in finite-feature density and normalized total persistence.} Panels (a,b) show the numbers of finite $\mathcal{H}_1$ and $\mathcal{H}_2$ features per 1,000 retained edges and triangles, respectively. Panels (c,d) show summed finite $\mathcal{H}_1$ and $\mathcal{H}_2$ lifetimes in years per 1,000 retained edges and triangles, respectively. Panels (e,f) show the numbers of finite $\mathcal{H}_1$ and $\mathcal{H}_2$ features with lifetimes of at least the indicated number of years, per 1,000 retained edges and triangles, respectively. Topological thresholding is shown in coral, marginal likelihood in verdigris, and the noise-corrected threshold in charcoal blue; the remaining eight methods are gray. Marginal likelihood has the highest alternative values for both aggregate $\mathcal{H}_1$ measures, and the noise-corrected threshold has the highest for both aggregate $\mathcal{H}_2$ measures. Retained edges enter each temporal filtration in the year when their endpoint concepts first co-occur. Topological thresholding uses the standard persistence image construction and constraint percentiles (50,25); statistical thresholds use $\alpha=0.05$ or the corresponding noise-corrected cutoff. Panels (a--d) include all finite lifetimes. Panels (e,f) show 0--40 years; fewer than 1.0\% of finite $\mathcal{H}_1$ features and 0.2\% of finite $\mathcal{H}_2$ features have lifetimes of at least 40 years in every network. Features that remain unfilled at the end of the filtration are excluded from these finite-feature measures.}
\sbox{\figTwelveBox}{\begin{minipage}{\textwidth}\fontsize{7.5}{9}\selectfont\figTwelveCaption\end{minipage}}
\begin{figure}[p]
\centering
\includegraphics[height=\dimexpr\textheight-\ht\figTwelveBox-\dp\figTwelveBox-\baselineskip\relax,keepaspectratio]{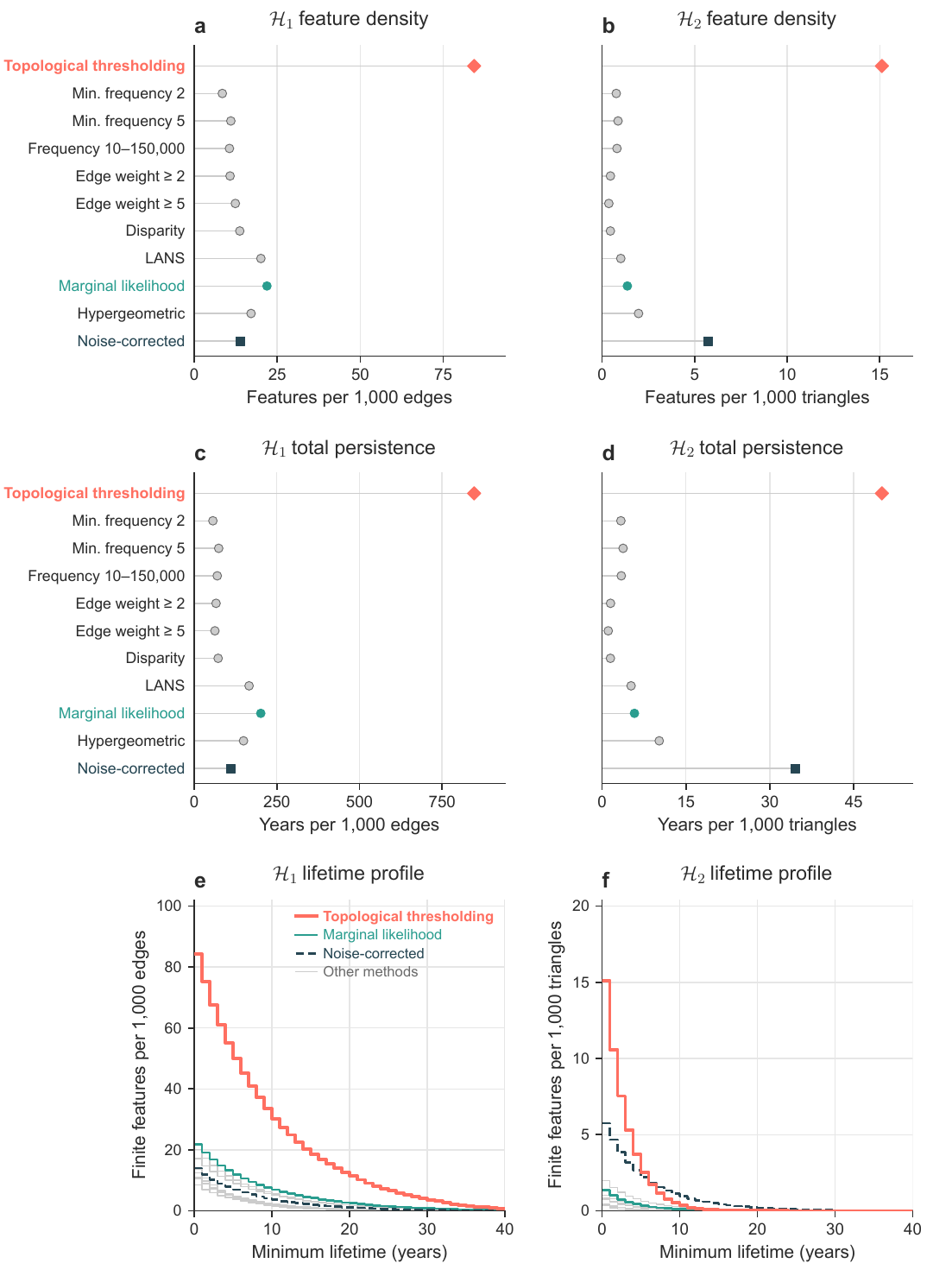}
\caption{\fontsize{7.5}{9}\selectfont\figTwelveCaption}
\label{fig:thresholding_comparison}
\end{figure}
\restoregeometry

\paragraph{Performance across Fields}
To assess performance beyond the applied mathematics example, we repeat the comparison using concept networks constructed from Dimensions article extracts for 154 scientific fields such as ecology, mechanical engineering, clinical sciences, and sociology. Within each field, we calculate document frequencies and co-occurrence counts, apply the eleven methods to the field's unthresholded network, and compute the same four measures. Topological thresholding uses the same constraint percentiles $(\delta_1,\delta_2)=(P_1(50),P_2(25))$ and the same candidate threshold grid expressed as fractions of the field's article count. The fixed frequency and edge count cutoffs and the statistical threshold settings remain unchanged across fields. Appendix~\ref{app:comparison_methods} describes the field extracts and network construction.

We summarize relative performance by comparing methods within each field before combining results across fields. For each alternative and measure, we divide the topological-thresholding value by the alternative's value, take the median of the logarithms of these paired ratios, and transform back to the ratio scale. Each eligible field contributes equally, and a ratio above one indicates a higher value for topological thresholding. Ratios require positive values for both methods; Table~\ref{tab:comparison_cross_field} reports the corresponding sample sizes and also gives paired win, tie, and loss counts that include defined zero outcomes. A measure with a zero denominator is undefined and is excluded from that comparison.

\begin{figure}[!htbp]
\centering
\includegraphics[width=\textwidth]{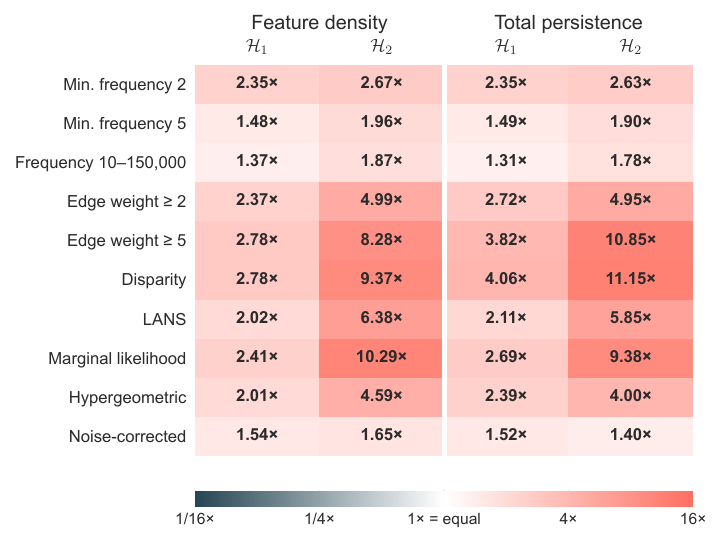}
\caption{\textbf{Across fields, median ratios favor topological thresholding over each alternative on all four measures.} For each measure, we divide the value for topological thresholding by the value for the row alternative within each field. Cells show medians across fields, calculated on the logarithmic ratio scale using positive values for both methods. Ratios above 1$\times$ indicate higher values for topological thresholding. Feature density counts finite $\mathcal{H}_1$ and $\mathcal{H}_2$ features per 1,000 retained edges and triangles, respectively. Normalized total persistence sums finite $\mathcal{H}_1$ and $\mathcal{H}_2$ lifetimes in years per 1,000 retained edges and triangles, respectively. Topological thresholding uses the selected persistence image construction and constraint percentiles $(\delta_1,\delta_2)=(P_1(50),P_2(25))$; statistical thresholds use $\alpha=0.05$ or the corresponding noise-corrected cutoff. Each cell summarizes 87--141 fields; Table~\ref{tab:comparison_cross_field} reports the sample size for each comparison and wins, ties, and losses, including defined zero values in those counts. Features that remain unfilled at the end of the filtration are excluded from these finite-feature measures.}
\label{fig:cross_field_comparison}
\end{figure}

Across fields, the median ratio favors topological thresholding over every alternative on all four measures (Figure~\ref{fig:cross_field_comparison}). Depending on the alternative, median feature-density ratios range from 1.37 to 2.78 for $\mathcal{H}_1$ and from 1.65 to 10.29 for $\mathcal{H}_2$. Median normalized-persistence ratios range from 1.31 to 4.06 for $\mathcal{H}_1$ and from 1.40 to 11.15 for $\mathcal{H}_2$. 

\subsection{Effects of Thresholding on the Concept Network}\label{subsec:application_effects}

The preceding comparison evaluates topological thresholding against ten alternative thresholding methods using finite-feature densities and normalized total persistence. For the concept-network application, the practical consequences also depend on how much the network shrinks and how the concepts it removes relate to those retained. We examine these changes in the applied mathematics network, distinguishing the contributions of the lower and upper frequency bounds and comparing the labels of removed and retained concepts.

At the selected constraint percentiles $(\delta_1,\delta_2)=(P_1(50),P_2(25))$, topological thresholding retains concepts appearing in 12--167 articles. The resulting network contains 7,979 of the 81,026 concepts and 210,930 of the 1,361,809 edges from the unthresholded network. These changes correspond to reductions of 90.2\% in concepts and 84.5\% in edges. Across the 128 selectable networks, concept retention ranges from 0.5\% to 39.3\% and edge retention from 0.01\% to 66.0\%; the selected network retains 9.8\% of concepts and 15.5\% of edges, placing it eighty-ninth highest in edge retention among the 128 selectable networks.

The lower and upper bounds remove different kinds of concepts. Applying the lower bound alone excludes the 72,692 concepts appearing in fewer than twelve articles and leaves 8,334 concepts and 431,574 edges. Applying the upper bound then removes a further 355 concepts and 220,644 edges, or 51.1\% of the edges remaining after the lower bound. The upper bound therefore removes few concepts but a large share of the links, because the concepts it excludes are the corpus's most frequent and most highly connected terms. The retained network has 11 components, with 7,968 of its 7,979 concepts in the largest component. Nine concepts are isolated, and the network’s median degree is 37.

Node removal can change how a scientific topic is represented without eliminating every concept related to that topic. For example, the removed high-frequency concept `differential equations' (9,559 articles) occurs within the retained `second order differential equations' (91 articles). Here, a general label is removed while a more specific concept within the same topic remains. Conversely, the removed low-frequency concept `active vehicle suspension system' (11 articles) contains the retained `suspension system' (153 articles). Comparing removed and retained labels therefore helps characterize which descriptions of related subject matter remain after thresholding.

We quantify lexical overlap through phrase containment, a relationship also considered in automatic term extraction \citep{frantzi2000automatic}. We count matches in both directions---whether the complete label of a removed concept occurs as a contiguous phrase within a retained label, and whether a retained label occurs within the removed label. A removed concept can satisfy both conditions.

Phrase overlap is more common among concepts removed by the upper frequency bound than among those removed by the lower bound. Of the 72,692 concepts below the lower bound, 24,688 (34.0\%) match at least one retained concept. Specifically, 23,429 contain a retained concept, 1,415 are contained in a retained concept, and 156 satisfy both conditions. Of the 355 concepts above the upper bound, 318 (89.6\%) match at least one retained concept. Here, 86 contain a retained concept, 293 are contained in a retained concept, and 61 satisfy both conditions. Among concepts with a match, low-frequency removals predominantly contain retained labels, whereas high-frequency removals predominantly occur within longer retained labels. 

\section{Discussion}\label{sec:discussion}

Network thresholding---pruning nodes or edges based on their properties---has become a standard preprocessing step, but existing methods suffer from fundamental limitations. Threshold selection typically relies on heuristics or trial and error, resulting in network structures that are highly sensitive to small parameter changes. Additionally, most methods consider only pairwise relationships, ignoring higher-order interactions among groups of three or more nodes that are increasingly recognized as pivotal to network structure and dynamics.

We address these limitations by introducing a novel algorithm that leverages topological data analysis to identify optimal threshold parameters. Our method uses persistent homology to encode higher-order network structures---including cycles, voids, and other topological features---and track how these structures evolve over a scale parameter such as distance, correlation, or time. By vectorizing persistence diagrams into persistence images, we represent each candidate network by a Euclidean vector indexed by its threshold parameters. The optimization routine favors stability of the network's topological representation under nearby threshold choices by minimizing the local change objective over all feasible grid points. User-specified constraints require minimum feature counts in each homological dimension considered, allowing analysts to tailor those requirements to their application.

We demonstrated this method on concept networks extracted from scientific literature, a common object of study in the Science of Science. In these networks, nodes represent scientific concepts and edges connect concepts that co-appear in article abstracts, capturing the conceptual landscape of a field and revealing how ideas relate and cluster. However, raw concept networks suffer from significant noise. Widely used terms create spurious connections that inflate network density without providing substantive insight, while very rare terms can fragment the network and complicate analysis. Effective thresholding based on concept frequency is therefore essential to filter this noise while preserving the underlying structure of scientific knowledge. We applied our algorithm to concept networks from applied mathematics, spanning publications from 1920 to 2021, where topological features have a natural interpretation as potential areas for conceptual innovation---cycles and cavities representing potential opportunities to `close gaps' by connecting previously unlinked concepts.

The comparisons support the use of higher-order structure in network threshold selection. Topological thresholding produces higher finite-feature densities and greater normalized total persistence than all ten alternatives in the applied mathematics network (Section~\ref{subsec:thresholding_comparison}). In dimension one, the larger persistence sum reflects both a greater density of finite features and a longer mean lifetime; in dimension two, it reflects density alone, since the mean lifetime is shorter than that of the strongest alternative. Across fields, the median paired comparison favors topological thresholding over each alternative on both measures in homological dimensions one and two, with constraint percentiles and alternative threshold settings held fixed.

\begin{figure}
    \centering
    \includegraphics[width=\linewidth]{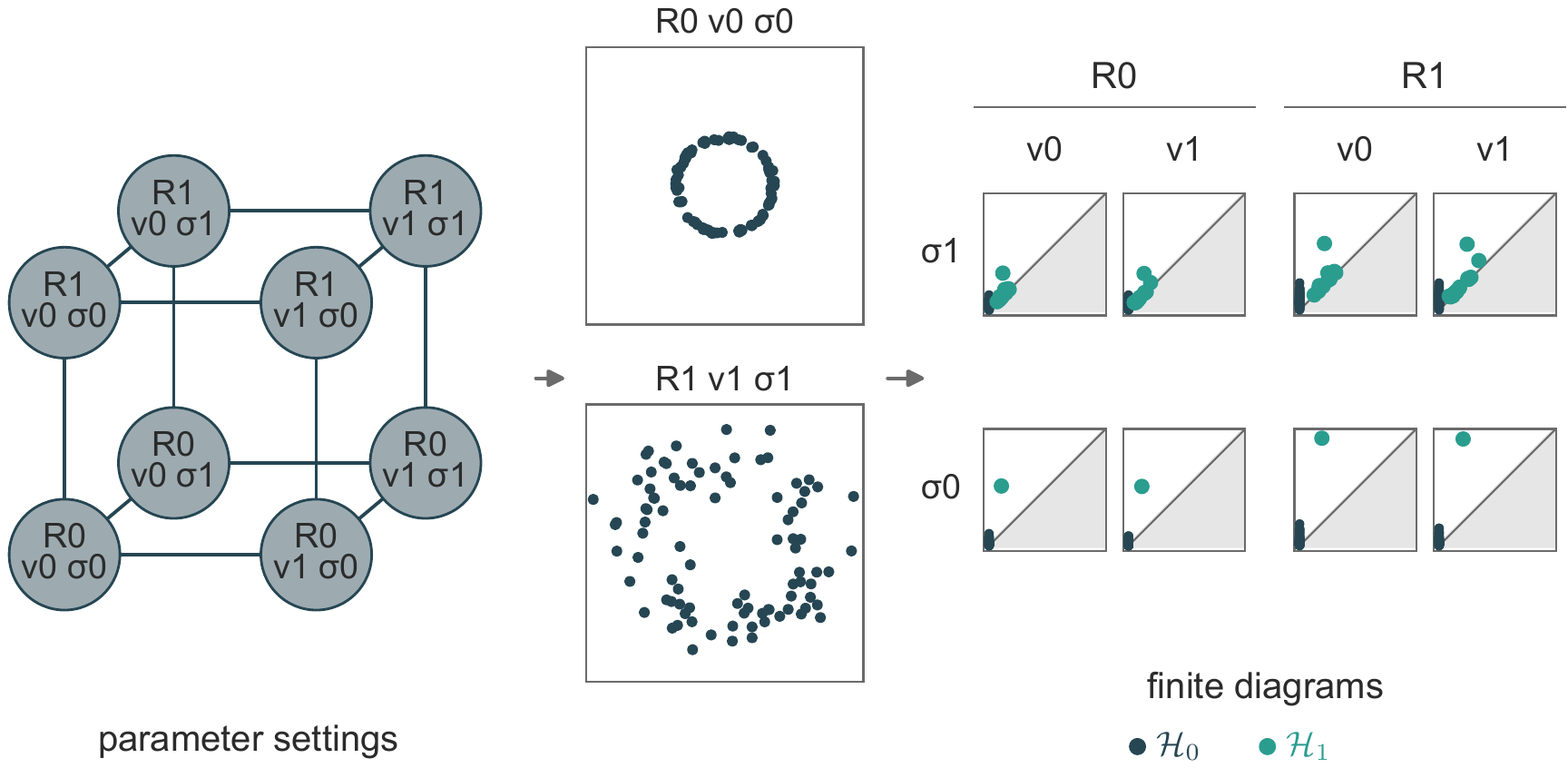}
    \caption{\textbf{A toy example of point cloud data over a three-dimensional parameter space.} The parameters are the radial parameter $R$ and the two noise parameters $\nu$ and $\sigma$. The pipeline shown on the top row of Figure \ref{fig:pipeline} maps the point cloud at each grid entry to its persistence diagram. The digits 0 and 1 label parameter indices. The diagrams form two $2\times2$ slices at fixed $R$, with $\nu$ indexing columns and $\sigma$ indexing rows. Birth is horizontal and death is vertical. Only finite $\mathcal{H}_0$ and $\mathcal{H}_1$ features are shown; one essential $\mathcal{H}_0$ class per cloud is omitted.}
    \label{fig:3d_domain}
\end{figure}

Our approach is not without limitations. The primary drawback of incorporating higher-order structure in persistent homology computations is increased computational complexity. Standard boundary-matrix reduction over a field has worst-case running time $\mathcal{O}(s^3)$, where $s$ is the number of simplices in the filtration \citep{zomorodian2009}. This bound describes the standard reduction algorithm; our computations use GUDHI's persistent-cohomology implementation \cite{gudhi:PersistentCohomology}. Our computation of $\mathcal{H}_1$ and $\mathcal{H}_2$ uses the flag (clique) complex through dimension three, including tetrahedra. For a graph with $v$ vertices, the number of simplices is therefore at most $\sum_{r=1}^{4}\binom{v}{r}=\mathcal{O}(v^4)$, with the actual number depending on the graph's clique structure. Practical runtimes can be substantially lower than the worst-case bound. Additionally, runtime scales with both the dimensionality and refinement of the parameter space. Doubling the resolution of a discretized two-dimensional parameter space results in four times the number of grid points, rapidly increasing computational cost. An alternative approach to our algorithm that may lower computation cost and could be considered in future research would be a local search over neighboring grid points using the local change objective of Equation \eqref{optimization_step} rather than an exhaustive evaluation of the grid. More research and validation would be needed to determine whether such an approach reaches the same selections as the grid search under the same constraints.

Although we have demonstrated threshold optimization in the specific context of dynamic networks, the algorithm is designed to be applicable to general data parameterization problems. For example, time series delay embeddings, sensor placements, and clustering algorithms all depend on one or more controllable parameters that could benefit from systematic optimization. Another potential application is point cloud data, as illustrated in Figure \ref{fig:3d_domain}. In this example, point cloud data is generated according to three parameters: a radial parameter $R$ and two noise parameters, $\nu$ and $\sigma$. The parameter space is therefore $\mathbb{R}^3$, and the persistence diagrams computed from the point clouds are indexed by a three-dimensional grid of these parameter values. In general, when the parameter domain is a sub-volume $U \subset \mathbb{R}^D$ for some positive integer $D$, the resulting persistence image vectors are indexed by a $D$-dimensional parameter grid.

\backmatter

\bmhead{Supplementary information} Not applicable




\bmhead{Acknowledgements} The authors acknowledge the Minnesota Supercomputing Institute at the University of Minnesota for providing resources that contributed to the research results reported within this paper. The authors also acknowledge Tom Gebhart (University of Minnesota) for insights which contributed to the manuscript.



\bmhead{List of abbreviations} ANZSRC, Australian and New Zealand Standard Research Classification; GUDHI, Geometry Understanding in Higher Dimensions; HONS, higher-order network selection; LANS, locally adaptive network sparsification; ML, machine learning; NSF, National Science Foundation; PD, persistence diagram; PI, persistence image; RREF, reduced row echelon form; TF-IDF, term frequency--inverse document frequency; VR, Vietoris--Rips

\bmhead{Availability of data and materials} We provide HONS (higher-order network selection), our Python implementation of topological thresholding, on GitHub (\url{https://github.com/rfunklab/hons}) and PyPI (\url{https://pypi.org/project/hons/}). Our companion replication repository (\url{https://github.com/rfunklab/hons-replication}) provides persistence diagrams, network-level measurements and scripts that reproduce the reported threshold selections, higher-order variance and method-comparison statistics from these derived inputs. The authors' code and derived numerical files are distributed under the MIT license.

The repository also includes a runnable example using open workplace contact observations from SocioPatterns (\url{https://sociopatterns.org/datasets/contacts-in-a-workplace/}), a research collaboration that measures face-to-face contacts using wearable sensors. This example constructs a network from contact records and applies topological thresholding to its edges. The workplace observations are distributed under CC0. The source Dimensions article--concept records used for the empirical analyses are proprietary and are not included in the repository.

\bmhead{Competing interests} The authors declare that they have no competing interests.

\bmhead{Funding} The authors thank the National Science Foundation (NSF) for financial support of work related to this project (grants BCS-2318171, CDS\&E-MSS-1854703, BCS-2318172, SMA-1829168, and SMA-1932596).

\bmhead{Authors' contributions} AS and JG devised the algorithm and prepared the original figures. AS derived the variance analysis and drafted the manuscript. RF led the additional analyses and preparation of the revised figures, software and replication materials. LZ, RF, and TO advised the research and contributed to and revised the manuscript.










\begin{appendices}
\renewcommand{\theHfigure}{appendix.\arabic{figure}}
\renewcommand{\theHtable}{appendix.\arabic{table}}
\renewcommand{\theHequation}{appendix.\arabic{equation}}

\section{Computing Homology on a Simplicial Complex}\label{secA3}

Given a simplicial complex, such as a Vietoris--Rips complex, homology groups can be computed as follows. We introduce an algebraic structure called a \emph{chain complex}, denoted $(C_k, \partial_k)_{k \in \mathbb{Z}}$, where $C_k$ is generated by the $k$-simplices and $\partial_k:C_k \to C_{k-1}$ is a homomorphism called the \emph{boundary operator} or \textit{boundary map} that reveals the boundaries of the $k$-simplices (which are constructed using $(k-1)$-simplices). More formally, an element of the $k$th chain group $C_k$ is a formal sum $\sum_i a_i \sigma_i$ of $k$-simplices using coefficients $a_i$, where $\sigma_i=[v_0...v_k]$. If $a_i$ come from a field, $C_k$ is a vector space; if $a_i$ come from a ring, then it is a module. The boundary operator, when applied to $\sigma=[v_0...v_k]$ returns $\sum_i (-1)^i (\hat{\sigma_i})$ where $\hat{\sigma_i}$ denotes the simplex $\sigma$ modified by omitting the $i$th $0$-simplex, $v_i$. Notably, the boundary operator satisfies $\partial_{k-1} \circ \partial_k =0$, i.e. the image in $C_{k-1}$ of the boundary operator $\partial_k$ is in the kernel of $\partial_{k-1}$.

The image of $\partial_{k+1}$ consists of boundaries of $(k+1)$-chains. We call the image the boundaries, denoted $\mathcal{B}_k$. The kernel of $\partial_k$ corresponds to cycles, denoted as $\mathcal{Z}_k$. As mentioned above, the image of $\partial_{k+1}$ is a subset of this kernel, but we may have some cycles that are not boundaries of any $(k+1)$-chains (these are `holes' in the data). Therefore, we can find the $k$th homology group as the quotient group
\begin{equation} \label{homology_k_group}
    \mathcal{H}_k = \mathcal{Z}_k / \mathcal{B}_k,
\end{equation}
whose elements are equivalence classes of cycles: two cycles represent the same class when their difference is a boundary.

\begin{figure}
    \centering
    \includegraphics[width=0.66\linewidth]{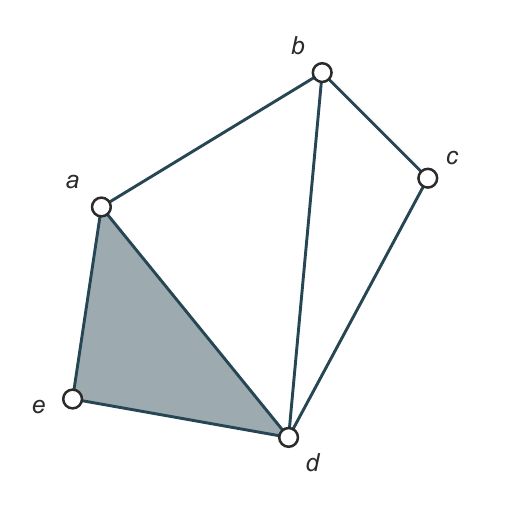}
    \caption{\textbf{A toy simplicial complex on vertices $a$, $b$, $c$, $d$, and $e$.} Notice that there are five $0$-simplices (the vertices themselves), seven $1$-simplices (all edges), but only one $2$-simplex, triangle $aed$, while $abd$ and $bcd$ are empty. We would intuitively expect to find that the dimension of $\mathcal{H}_1$ is $2$, corresponding to the number of planar cycles.}
    \label{fig:appendix_simplicial_complex_example}
\end{figure}

To demonstrate this principle, consider the toy simplicial complex $\Sigma$ shown in Figure \ref{fig:appendix_simplicial_complex_example}. There is only one $2$-simplex, triangle $ade$. In the matrices below, RREF denotes reduced row echelon form. Using modulo two integer coefficients $\mathbb{Z}_2$, we can represent the effect of the homomorphism $\partial_2$ as the linear map
\begin{align}\label{b_2_map}
\partial_2(\Sigma)=
\begin{blockarray}{cc}
ade \\
\begin{block}{(c)c}
  1 & ad \\
  1 & de \\
  1 & ea \\
\end{block}
\end{blockarray}
\overset{\text{RREF}}{\longrightarrow}
\begin{blockarray}{cc}
ade \\
\begin{block}{(c)c}
  1 & ad \\
  0 & de \\
  0 & ea \\
\end{block}
\end{blockarray}
\end{align}
Row reduction shows that $\operatorname{im}\partial_2$ has dimension one, so $\mathcal{B}_1$ has dimension one. Since $ade$ is the only $2$-simplex, this boundary matrix has one column. Zero rows are omitted from the displayed $\partial_2$ matrix. The boundary matrix for $\partial_1$ includes all seven edges:
\begin{align}\label{b_1_map}
\begin{split}
\partial_1(\Sigma) =
&\begin{blockarray}{cccccccc}
ab & bc & cd & bd & da & de & ea\\
\begin{block}{(ccccccc)c}
  1 & 0 & 0 & 0 & 1 & 0 & 1 & a \\
  1 & 1 & 0 & 1 & 0 & 0 & 0 & b \\
  0 & 1 & 1 & 0 & 0 & 0 & 0 & c \\
  0 & 0 & 1 & 1 & 1 & 1 & 0 & d \\
  0 & 0 & 0 & 0 & 0 & 1 & 1 & e \\
\end{block}
\end{blockarray} \\
\overset{\text{RREF}}{\longrightarrow}
&\begin{blockarray}{cccccccc}
ab & bc & cd & bd & da & de & ea\\
\begin{block}{(ccccccc)c}
  1 & 0 & 0 & 0 & 1 & 0 & 1 & a \\
  0 & 1 & 0 & 1 & 1 & 0 & 1 & b \\
  0 & 0 & 1 & 1 & 1 & 0 & 1 & c \\
  0 & 0 & 0 & 0 & 0 & 1 & 1 & d \\
  0 & 0 & 0 & 0 & 0 & 0 & 0 & e \\
\end{block}
\end{blockarray}
\end{split}
\end{align}
The kernel of $\partial_1$ has dimension $3$, so $\mathcal{Z}_1$ has dimension $3$. Over $\mathbb{Z}_2$, the quotient therefore has dimension $\dim \mathcal{H}_1=\dim \mathcal{Z}_1-\dim \mathcal{B}_1=3-1=2$, consistent with the two unfilled cycles in the figure.

\section{Parameter Assignments and Hyperparameter Tuning}\label{secA2}

For the 128 selectable applied mathematics networks, the lower bound $\ell$ assumes values in $\{3.3,5.0,6.7,8.4,10.0,11.7,13.4,15.0\}$, and the upper bound $u$ assumes values in $\{16.7,33.4,50.1,66.8,83.5,125.3,167.0,334.0,501.0,668.0,835.0,1{,}002.0,\linebreak1{,}169.0,1{,}336.0,1{,}503.0,167{,}001\}$. These numerical values were generated using percentages of the total number of papers in the applied mathematics corpus, resulting in non-integral parameterizations. For the lower bounds, these percentages were $\{0.002,0.003,0.004,0.005,0.006,0.007,0.008,0.009\}$, while for the upper bounds these percentages were $\{0.01,0.02,0.03,0.04,0.05,\linebreak0.075,0.1,0.2,0.3,0.4,0.5,0.6,0.7,0.8,0.9,100\}$. The displayed bounds are rounded; concept retention uses the ceilings of the unrounded lower bounds and the floors of the unrounded upper bounds. The full grid additionally includes $\ell=1.67001$ (0.001\% of the corpus) as a neighboring column for the local change objective, giving 144 grid entries in total. 

Table \ref{tab:hyperparameters} lists the ten percentile settings for each constraint used in Section \ref{subsec:results}. We evaluate all 100 pairs in their Cartesian product.

\begin{sidewaystable}
    \centering
    \caption{Percentile settings for the feature-count constraints.}\label{tab:hyperparameters}
   \begin{tabular}{lcccccccccc}
   \toprule
   Dimension $k$ & \multicolumn{10}{c}{Constraint value $\delta_k$}\\
   \midrule
    1 & $P_1(1)$ & $P_1(11)$ & $P_1(21)$ & $P_1(31)$ & $P_1(41)$ & $P_1(51)$ & $P_1(61)$ & $P_1(71)$ & $P_1(81)$ & $P_1(91)$\\
    2 & $P_2(1)$ & $P_2(11)$ & $P_2(21)$ & $P_2(31)$ & $P_2(41)$ & $P_2(51)$ & $P_2(61)$ & $P_2(71)$ & $P_2(81)$ & $P_2(91)$\\
    \botrule
    \end{tabular}
    \par\smallskip
    \begin{minipage}{\linewidth}\footnotesize
    \emph{Notes:} The application in Section \ref{subsec:results} uses all 100 combinations in the Cartesian product of the dimension-one and dimension-two arrays. $P_k(q)$ denotes the $q$th percentile of the number of $k$-dimensional homological features taken over the 128 selectable networks of the grid; the sweep uses the 1st, 11th, \ldots, 91st percentiles in each dimension.
    \end{minipage}
\end{sidewaystable}

\subsection{Persistence Image Bandwidth and Resolution}\label{subsec:image_sensitivity}

The adopted persistence images use a Gaussian standard deviation of $\sigma=0.1$ on a $20\times20$ grid. To assess how the optimal selections depend on these two choices, we recompute them over the Cartesian product of $\sigma\in\{0.05,0.1,0.2\}$ and resolutions $10\times10$, $20\times20$ and $40\times40$, holding fixed the weighting function, the PI domain, the treatment of essential classes, the set of selectable candidates, and the percentile convention for the constraints. All nine configurations use the persistence diagrams from the same 144 grid networks. Because the constraints are percentiles of the feature count distributions, and those counts do not depend on the PI construction, the constraint thresholds and the feasible set are identical at every configuration.

Table \ref{tab:image_sensitivity} compares each configuration with the adopted one at the same constraint setting, over the 99 of 100 sweep settings that admit a feasible network. Agreement is 96 of 99 at $\sigma=0.05$ and 98 of 99 at $\sigma=0.2$, and selections were unchanged across the three resolutions tested. No constraint setting produced more than one optimal network at any configuration. The four constraint settings with differing selections all use the first-percentile requirement in dimension one, $\delta_1=P_1(1)$, where that constraint admits nearly the whole grid; no selection differs at a stricter requirement. For each of the three illustrated constraint settings of Figure \ref{fig:fig_2_reworked}, the selected network is unchanged across all nine PI configurations.

\begin{table}[h]
    \centering
    \caption{Selection agreement across persistence image bandwidths and resolutions.}\label{tab:image_sensitivity}
    \begin{tabular}{cccccc}
    \toprule
    $\sigma$ & Resolution & Same selection & Different selection & Agreement & Settings with ties\\
    \midrule
    0.05 & $10\times10$ & 96 & 3 & 97.0\% & 0\\
    0.05 & $20\times20$ & 96 & 3 & 97.0\% & 0\\
    0.05 & $40\times40$ & 96 & 3 & 97.0\% & 0\\
    0.10 & $10\times10$ & 99 & 0 & 100.0\% & 0\\
    0.10 & $20\times20$ & 99 & 0 & 100.0\% & 0\\
    0.10 & $40\times40$ & 99 & 0 & 100.0\% & 0\\
    0.20 & $10\times10$ & 98 & 1 & 99.0\% & 0\\
    0.20 & $20\times20$ & 98 & 1 & 99.0\% & 0\\
    0.20 & $40\times40$ & 98 & 1 & 99.0\% & 0\\
    \botrule
    \end{tabular}
    \par\smallskip
    \begin{minipage}{\linewidth}\footnotesize
    \emph{Notes:} Each of the nine combinations of Gaussian standard deviation and image resolution is compared with the adopted configuration ($\sigma=0.1$, $20\times20$) at the same constraint setting. Counts run over the 99 sweep settings that admit a feasible network; $(P_1(91),P_2(91))$ admits none and is excluded. The last column counts settings at which more than one feasible network attains the minimum objective.
    \end{minipage}
\end{table}

\section{Threshold Optimization over Zoology Networks}\label{SecA4}

\begin{figure}
    \centering
    \includegraphics[width=0.74\linewidth]{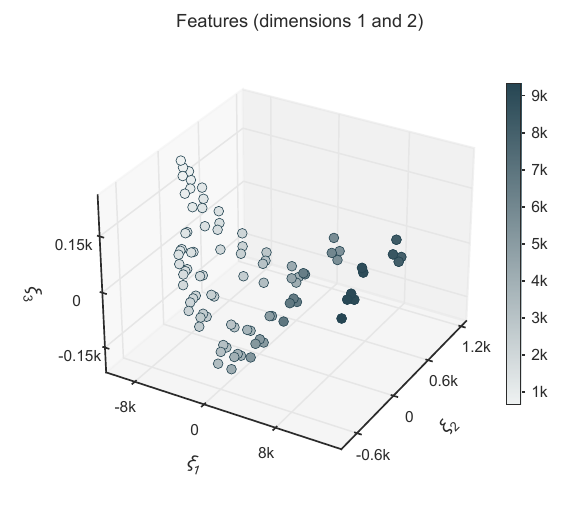}
    \caption{\textbf{PI vectors for the zoology dataset.} The vectors over the grid specified in Appendix~\ref{SecA4} are projected onto the first three principal components $(\xi_1, \xi_2, \xi_3)$ and colored according to the sum of the dimension one and two feature counts.}
    \label{fig:zoology_manifold}
\end{figure}

\begin{figure}
    \centering
    \includegraphics[width=\linewidth]{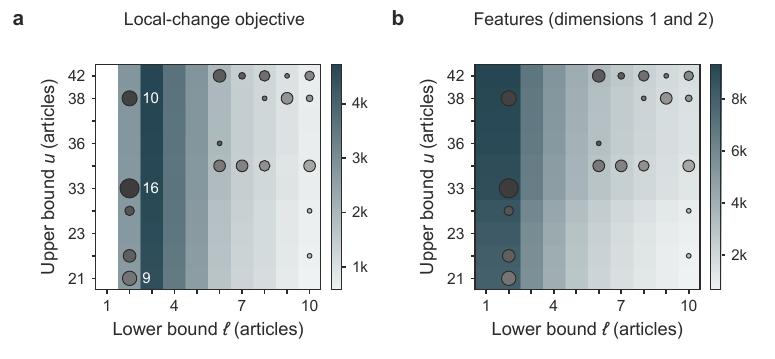}
    \caption{\textbf{All 100 constraint settings are feasible and select 20 distinct grid cells in zoology.} Selections over 100 percentile combinations of $\delta_1$ and $\delta_2$ are shown over (a) the local change objective and (b) the dimension one and two feature counts. Each bubble sits at a selected grid cell; its area is the number of constraint settings that select that cell and its shade is the mean sum of the two constraint percentiles across those settings, with lighter tones for looser constraints, as in Figure \ref{fig:varied_tuning_parameters}. Numbers label the selection counts for the three most frequently selected cells. The objective and the selections are defined on the 90 selectable networks; the $\ell=1$ column is built because the local change objective uses it as a neighbor, and is never selectable.}
    \label{fig:zoology_results}
\end{figure}

In this section, we demonstrate a second application of the pipeline to a smaller dataset, namely concept networks describing the evolution of the academic field of zoology. We generate a network at each of 100 unique parameterizations of upper bound $u$ (maximum number of articles in which a concept may appear) and lower bound $\ell$ (minimum number of articles in which a concept must appear). The upper bound assumes values in $\{21,22,23,25,33,34,36,37,38,42\}$, while the lower bound assumes values in $\{1,2,3,4,5,6,7,8,9,10\}$. These networks use the same preparation and the same persistence image construction as the applied mathematics example. As there, the $\ell=1$ column is built because the local change objective uses it as a neighbor but is never selectable, leaving 90 selectable networks; the componentwise mean in Equation \eqref{h_o_var} is taken over all 100 grid images. The resulting networks range from 672 to 9,332 dimension one and two features. The resulting PI vectors, projected onto the first three principal components, are shown in Figure \ref{fig:zoology_manifold}. After computing the local change objective, we track the optimal selection over the range of hyperparameter tunings described in Appendix \ref{secA2}, as in Section \ref{subsec:results}. Here $P_k(q)$ is computed over the 90 selectable zoology networks. Figure \ref{fig:zoology_results} illustrates the optimal selection's dependence on hyperparameter tuning: all 100 settings admit a feasible cell and they concentrate on 20 distinct cells, with one cell ($u=33$, $\ell=2$) selected by 16 of them.

\begin{figure}
    \centering
    \includegraphics[width=0.755\linewidth]{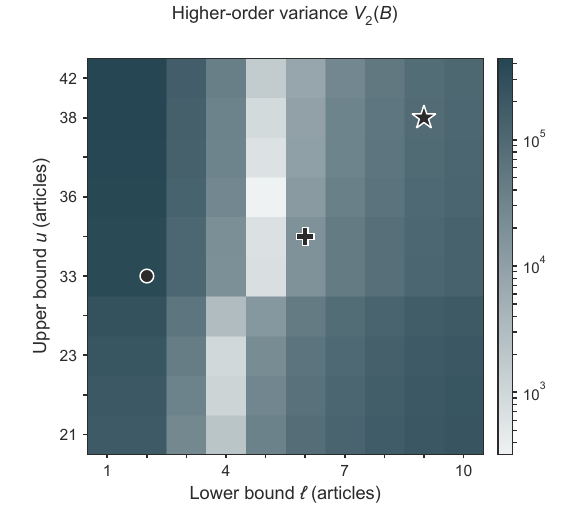}
    \caption{\textbf{Higher-order variance over the zoology networks.} Values are evaluated with Equation \eqref{h_o_var}. The componentwise mean is taken over all 100 grid images; the ranks below are among the 90 selectable networks. Markers show the three highlighted selections, as in Figure \ref{fig:fig_2_reworked}. The selection at $(P_1(50),P_2(25))$ (plus sign, $u=34$, $\ell=6$) has the 16th-lowest variance of the 90, and the 14th-lowest of the 39 networks satisfying those constraints. The selections at $(P_1(25),P_2(25))$ (star, $u=38$, $\ell=9$) and $(P_1(75),P_2(50))$ (dot, $u=33$, $\ell=2$) rank 49th and 86th of the 90.}
    \label{fig:zoology_variance}
\end{figure}

We likewise use higher-order variance to examine whether the locally stable selection is close to the grid average in its relative representation of cycles and voids. The heatmap of variance measures (taken according to Equation \eqref{h_o_var}) is shown in Figure \ref{fig:zoology_variance}. Under the primary constraints $(\delta_1,\delta_2)=(P_1(50),P_2(25))$ the selection has the 16th-lowest variance among the 90 selectable networks, and the 14th-lowest among the 39 that satisfy those constraints; the minimum over the grid lies at $u=36$, $\ell=5$. The selections at $(P_1(25),P_2(25))$ and $(P_1(75),P_2(50))$ rank 49th and 86th of the 90.

\section{Notes on the Higher-Order Variance}\label{sec:notes_on_variance}

This appendix discusses the motivation for Equation \eqref{h_o_var} and an alternative measure that extends to additional homological dimensions.

Let $\hat{\rho}_{B_k}=\rho_{B_k}-\overline{\rho_k}$ be the departure of network $B$'s persistence image in homological dimension $k$ from its componentwise grid mean, as defined in Section~\ref{sec:statistical_implications}. Equation~\eqref{h_o_var} can be written as
\begin{equation}\label{variance_of_difference}
\begin{aligned}
V_2(B)&=\frac{\|\hat{\rho}_{B_1}\|_2^2+\|\hat{\rho}_{B_2}\|_2^2-2\hat{\rho}_{B_1}^{T}\hat{\rho}_{B_2}}{n-1}\\
&=\frac{\|\hat{\rho}_{B_1}-\hat{\rho}_{B_2}\|_2^2}{n-1}.
\end{aligned}
\end{equation}
The first two terms measure each PI's squared departure from its grid mean. The cross term lowers the statistic when those departures align at corresponding pixels and raises it when they oppose one another. Higher-order variance therefore measures the departure of the difference between the two PIs from its grid average. The factor $1/(\eta-1)$ sets the scale for PIs with $\eta$ pixels.

If the aim is instead to summarize each homological dimension's departure from its grid mean separately, we can use a scaled sum of squared distances.
\begin{equation}\label{alt_h_o_var}
V_{k_{\text{max}}}^p(B) := \frac{\alpha_p}{k_{\text{max}}} \sum_{k=1}^{k_{\text{max}}} ||\rho_{B_k} - \overline{\rho_k}||_p^2,
\end{equation}
where $p$ and $k_{\text{max}}$ are as defined in Section \ref{mathematical_details_algorithm}, and $\alpha_p$ is a scale parameter depending on the choice of norm (e.g. for Euclidean norms, $\alpha_p = (\eta-1)^{-1}$). For $p=2$, this measure averages the squared departures of the individual images from their grid means, with scale factor $\alpha_p$. Other norms allow the measure to be tailored to a particular application.





\section{Comparison Methods}\label{app:comparison_methods}

Section~\ref{subsec:thresholding_comparison} compares topological thresholding with ten methods that select nodes or edges using document frequencies, co-occurrence counts, or statistical criteria. The comparison first examines the applied mathematics network developed in the main analysis, then evaluates performance across scientific fields. This appendix defines the input networks and thresholding rules, followed by the temporal construction and summary measures used to compare the resulting networks.

\subsection{Input Networks and Threshold Selection}

Every method starts from the same input network within a corpus. The applied mathematics corpus described in Section~\ref{subsec:data_generation} contains 167,001 articles published from 1920 to 2021, with the final year only partially observed. The corresponding unthresholded network has 81,026 concept nodes and 1,361,809 undirected edges. For each concept $i$, document frequency $d_i$ counts the distinct articles containing that concept. An edge joins two concepts that occur together in at least one article, and its co-occurrence count $c_{ij}$ is the number of distinct articles containing both. These frequencies and counts are calculated before any of the comparison thresholds are applied.

The detailed applied mathematics input uses the same preparation as the cross-field evaluation. We retain concept records from 1920--2021 with relevance scores of at least 0.7, deduplicate article--concept pairs, and then retain concepts appearing in at least two articles of the resulting corpus. No cap is placed on the number of concepts an article may contribute, and document frequencies are counted after these restrictions, as is the article count $N$ that scales the threshold grid. The detailed example and the cross-field evaluation therefore differ only in which field-of-research code they draw on, not in how the corpus is prepared.

The cross-field comparison uses separate Dimensions article--concept tables for 154 ANZSRC field groups. We include article--concept records from publication years 1920--2021 with a concept relevance score of at least 0.7. After applying these criteria, we retain concepts occurring in at least two distinct articles within a field and remove duplicate article--concept pairs. The corpus size $N$ counts distinct articles remaining in the field's table; articles with no retained concepts are therefore absent from $N$. An article assigned to multiple fields can contribute to each field.

Network measurements were completed for 152 of the 154 fields. One field had no co-occurrence edges, and Geology exceeded the implementation's simplex-key limit. Ten of the 152 measured fields had no edges after topological thresholding, leaving 142 fields with at least one defined paired comparison. Table~\ref{tab:comparison_cross_field} reports the eligible sample for each measure and alternative.

Topological thresholding selects a document-frequency interval using the procedure in the main text, with constraint percentiles fixed at $(\delta_1,\delta_2) = (P_1(50), P_2(25))$. For a corpus of $N$ articles, the lower candidate bounds are $N\times\{1,2,\ldots,9\}\times10^{-5}$ and the upper bounds are
\[
N\times\{1,2,3,4,5,7.5,10,20,30,40,50,60,70,80,90,10{,}000\}\times10^{-4}.
\]
The resulting grid contains 144 candidate networks. The eight larger lower bounds give the 128 networks eligible for selection; the smallest lower bound supplies a neighboring network for the finite-difference calculation. A concept is retained when $\lceil\ell\rceil\leq d_i\leq\lfloor u\rfloor$. The percentile constraints are computed over the selectable networks, and the optimal network minimizes the local change objective among those meeting both constraints. The feature counts used for selection include classes that remain unfilled at the end of the filtration. The finite feature measures below are calculated after selection and do not replace those constraints.

\subsection{Fixed Node and Edge Thresholds}

Fixed node based thresholds compare the interval selected by topological thresholding with simple cutoffs on the same scalar property. We retain concepts appearing in at least two articles, concepts appearing in at least five articles, or concepts within the interval of 10--150,000 articles. In the detailed applied mathematics example, the maximum frequency is 9,559, so the upper bound of 150,000 excludes no concepts. Because the cross-field input tables already exclude concepts observed in only one article, the minimum-frequency-two rule retains every input node in that comparison. All three fixed rules retain the subgraph induced by the selected nodes. Topological thresholding uses the same induced subgraph convention: an edge survives when both endpoints are retained, and qualifying nodes remain as vertices even if they have no surviving edges.

Global edge weight thresholds retain pairs of concepts that co-occur repeatedly. We retain an edge when $c_{ij}\geq2$ or $c_{ij}\geq5$. The count used for selection is distinct from the first co-occurrence year used to time the edge in the temporal filtration. An edge based output contains the retained edges and their endpoints, so a node with no surviving edge is absent. This vertex convention also applies to the statistical thresholds below.

\subsection{Statistical Backbone Thresholds}

Statistical backbone thresholds compare an edge's observed weight with local weight distributions or a null expectation. Let $k_i$ be the degree of node $i$ in the unthresholded network, $s_i=\sum_jc_{ij}$ its strength, and $W=\sum_{i<j}c_{ij}$ the total co-occurrence weight, counting each undirected edge once. All scores use these unthresholded network quantities. Disparity, marginal likelihood, and hypergeometric thresholding use a per edge probability cutoff of $\alpha=0.05$, without multiple testing correction. LANS uses the same cutoff on an empirical tail fraction. The noise-corrected threshold uses the standard deviation multiplier specified below.

\subsubsection{Disparity Threshold}
The disparity threshold compares the share of a node's strength carried by an edge with a random partition of that strength among its incident edges~\cite{serrano2009extracting}. For an endpoint with $k_i>1$, the endpoint probability is
\[
p_{ij}^{(i)}=\left(1-\frac{c_{ij}}{s_i}\right)^{k_i-1}.
\]
An edge is retained when the probability is at most $\alpha$ at either scored endpoint. Degree one endpoints are not scored under the implementation used here. An edge with one degree one endpoint must therefore qualify through the other endpoint; an edge joining two degree one nodes is excluded.

\subsubsection{Locally Adaptive Network Sparsification}
LANS compares an edge with the empirical distribution of incident weights at each endpoint~\cite{foti2011}. Its endpoint score is the fraction of incident edges whose counts are strictly greater than the focal count,
\[
q_{ij}^{(i)}=\frac{|\{h:c_{ih}>c_{ij}\}|}{k_i}.
\]
We retain an edge when $q_{ij}^{(i)}\leq\alpha$ or $q_{ij}^{(j)}\leq\alpha$. Equal weights remain tied, and equality at the cutoff qualifies. At $\alpha=0.05=1/20$, the endpoint decision is evaluated as the exact integer inequality $20|\{h:c_{ih}>c_{ij}\}|\leq k_i$.

\subsubsection{Marginal Likelihood Threshold}
The marginal likelihood threshold uses endpoint strengths to define a binomial null distribution for an edge's co-occurrence count~\cite{dianati2016}. We use
\[
X_{ij}\sim\operatorname{Binomial}\!\left(W,\frac{s_i s_j}{2W^2}\right)
\]
and retain an edge when $\Pr(X_{ij}\geq c_{ij})\leq\alpha$. The inclusive upper tail counts an outcome equal to the observed weight as at least as extreme.

\subsubsection{Noise Corrected Threshold}
The noise corrected threshold evaluates the excess of an observed count over its expectation relative to the uncertainty in that excess~\cite{coscia2017}. We use the reference variance based rule with $n=2W$, $\mu_{ij}=s_i s_j/n$, and score
\[
b_{ij}=\frac{c_{ij}-\mu_{ij}}{c_{ij}+\mu_{ij}}.
\]
An edge is retained when $b_{ij}>z\sigma_{ij}$, where $\sigma_{ij}$ is the estimated score standard deviation and $z=\Phi^{-1}(0.95)$. Here $\Phi$ is the standard normal cumulative distribution function. Calculations use the unrounded value of $z$, giving an approximate correspondence with the probability cutoff used for the other thresholds.

We estimate the score variance using the beta--binomial calculation in the reference implementation of noise-corrected thresholding (\url{https://www.michelecoscia.com/?page_id=287}).

\subsubsection{Hypergeometric Threshold}
The hypergeometric threshold compares the observed overlap between two concepts with random assignment to articles while holding their document frequencies fixed~\cite{tumminello2011}. For a corpus of $N$ articles, with concepts present in $d_i$ and $d_j$ articles, the null overlap is $X_{ij}\sim\operatorname{Hypergeom}(N,d_i,d_j)$. We retain an edge when $\Pr(X_{ij}\geq c_{ij})\leq\alpha$. The calculation uses the full article set as the overlap population and does not stratify articles by the number of concepts they contain.

\subsection{Temporal Filtration and Feature Measurement}

Persistent homology evaluates the thresholded networks as co-occurrence links accumulate over publication time, following the temporal construction used in the main analysis. The thresholding first selects nodes or edges using counts from the full corpus. Each retained edge then enters the filtration at its first co-occurrence year, even if the count required for its selection is reached later. Thus, thresholding determines the final graph whose temporal development is evaluated; the frequency cutoffs are not reapplied year by year.

We map publication years to filtration coordinates and convert measured lifetimes back to calendar years. All retained vertices enter at filtration value zero. For every retained network, in the detailed applied mathematics comparison and in the cross-field comparison alike, an edge first observed in year $y$ enters at $(y-y_{\min})/(y_{\max}-y_{\min})$, where $y_{\min}$ is the earliest article year associated with a retained concept in that network and $y_{\max}$ is the latest article year in the field's input table. The span is therefore a property of each retained network rather than a fixed constant. In the detailed example, the topologically thresholded network spans 98 years (1923--2021), the two minimum-frequency alternatives span 101 years, and the remaining alternatives span 100 years. When the two years coincide, all filtration values are zero. A finite interval $[b,d)$ therefore has lifetime $(y_{\max}-y_{\min})(d-b)$ years in both comparisons, using that network's own span. Reporting calendar year differences makes the lifetime unit consistent across the two calculations.

The filtration is a flag complex built from the retained graph, a Vietoris-Rips complex. Each complete subgraph on three vertices contributes a triangle and each complete subgraph on four vertices contributes a tetrahedron. A simplex enters when its last edge appears; the corresponding concepts need not all co-occur in a single article. Including tetrahedra permits two-dimensional features to be filled. We compute $\mathcal{H}_1$ and $\mathcal{H}_2$ over $\mathbb{Z}/11\mathbb{Z}$ using GUDHI~\cite{gudhi_simplextree_comparison}. 

The comparison measures use finite persistence intervals with positive lifetime. For dimension $k\in\{1,2\}$, let $n_k$ be their count and $L_{kr}$ the calendar year lifetime of interval $r$. Zero length intervals and classes that remain unfilled at the end of the observation period are excluded. Counts describe the features occurring in each thresholded network.

We normalize feature counts and lifetime sums by the structural elements that form features in the corresponding dimension. Let $E$ and $T$ be the numbers of edges and triangles in the final retained graph, and set $m_1=E$ and $m_2=T$. The four aggregate measures are
\[
D_k=\frac{1{,}000\,n_k}{m_k},\qquad
S_k=\frac{1{,}000}{m_k}\sum_{r=1}^{n_k}L_{kr},\qquad k\in\{1,2\}.
\]
Here $D_k$ is finite feature density, and $S_k$ is normalized total finite persistence, in years per 1,000 edges or triangles. Triangles are counted once in the final undirected graph. With edges but no finite one-dimensional features, $D_1=S_1=0$; with triangles but no finite two-dimensional features, $D_2=S_2=0$. The measures are undefined when their edge or triangle denominator is zero.

The lifetime profiles show how feature density changes when short lived features are omitted. At minimum lifetime $t$, the plotted value is
\[
D_k(t)=\frac{1{,}000}{m_k}\sum_{r=1}^{n_k}\mathbf{1}\{L_{kr}\geq t\}.
\]
We evaluate these profiles using differences between the integer calendar years of birth and death, including intervals exactly at the cutoff. Figure~\ref{fig:thresholding_comparison} displays minimum lifetimes from 0 to 40 years. The aggregate counts and lifetime sums include all positive finite lifetimes.

\subsection{Summaries across Fields}

The cross-field comparison evaluates each alternative separately using paired measurements within fields. For a measure $M\in\{D_1,D_2,S_1,S_2\}$, let $M_f^{\mathrm{TT}}$ and $M_f^{a}$ be its values for topological thresholding and alternative $a$ in field $f$. For fields with both values positive, define $r_{fa}=M_f^{\mathrm{TT}}/M_f^a$. The displayed summary is
\[
R_a=2^{\operatorname{median}_f(\log_2 r_{fa})}.
\]
Each eligible field has equal weight. With an even number of observations, this calculation gives the geometric midpoint of the two central ratios. The lower and upper quartiles are calculated on the same logarithmic scale using linear interpolation and transformed back to ratios.

Paired win, tie, and loss counts supplement the ratios by retaining comparisons with defined zero outcomes. A win means that the topological-thresholding value exceeds the alternative's value; a loss means the reverse. We classify values as tied when $|M_f^{\mathrm{TT}}-M_f^a|\leq10^{-12}+10^{-9}|M_f^a|$. Undefined measures and unavailable measurements are omitted from the corresponding pair rather than recorded as zeros. Table~\ref{tab:comparison_cross_field} reports both the number of positive pairs used for ratios and the number of defined pairs used for wins, ties, and losses.

\section{Supplementary Comparison Results}\label{app:comparison_results}
\setcounter{table}{0}
\renewcommand{\thetable}{F\arabic{table}}
\renewcommand{\theHtable}{comparison.\arabic{table}}

The two tables provide the numerical results underlying the comparisons in Section~\ref{subsec:thresholding_comparison}. Table~\ref{tab:comparison_applied_math} reports the retained graph sizes, finite-feature counts, densities, and normalized lifetime sums for the applied mathematics example. Table~\ref{tab:comparison_cross_field} summarizes each alternative's comparison with topological thresholding across fields, including variation among fields and the number of eligible pairs. Appendix~\ref{app:comparison_methods} defines all thresholding rules and measures.

\begin{table}[!htbp]
\centering
\caption{Network sizes and finite-feature measures in applied mathematics.}\label{tab:comparison_applied_math}
{\small\setlength{\tabcolsep}{4pt}\renewcommand{\arraystretch}{1.15}
\begin{tabular}{lrrrrr}
\toprule
\multicolumn{6}{l}{\textbf{(a) Retained graph and finite-feature counts}}\\[3pt]
Method & Nodes & Edges & Triangles & Finite $\mathcal{H}_1$ & Finite $\mathcal{H}_2$ \\
\midrule
Topological thresholding & 7\,979 & 210\,930 & 739\,045 & 17\,774 & 11\,172 \\
Min. frequency 2 & 81\,026 & 1\,361\,809 & 12\,064\,149 & 11\,534 & 9\,291 \\
Min. frequency 5 & 23\,982 & 780\,500 & 9\,565\,842 & 8\,608 & 8\,285 \\
Frequency 10--150,000 & 10\,321 & 491\,316 & 7\,720\,132 & 5\,215 & 6\,238 \\
Edge weight $\geq$ 2 & 46\,754 & 220\,724 & 1\,210\,390 & 2\,384 & 551 \\
Edge weight $\geq$ 5 & 10\,065 & 31\,397 & 105\,952 & 389 & 39 \\
Disparity & 13\,134 & 34\,834 & 88\,520 & 478 & 40 \\
LANS & 80\,040 & 603\,608 & 1\,410\,894 & 12\,119 & 1\,430 \\
Marginal likelihood & 80\,040 & 1\,107\,447 & 3\,570\,057 & 24\,255 & 4\,888 \\
Hypergeometric & 80\,004 & 1\,184\,425 & 5\,126\,146 & 20\,270 & 10\,115 \\
Noise-corrected & 80\,040 & 1\,290\,841 & 6\,682\,565 & 18\,089 & 38\,400 \\
\bottomrule
\end{tabular}
\par\medskip
\begin{tabular}{lrrrr}
\toprule
\multicolumn{5}{l}{\textbf{(b) Feature density and normalized total finite persistence}}\\[3pt]
 & \multicolumn{2}{c}{Feature density} & \multicolumn{2}{c}{\shortstack{Normalized total\\finite persistence}} \\
\cmidrule(lr){2-3}\cmidrule(lr){4-5}
Method & $\mathcal{H}_1$ & $\mathcal{H}_2$ & $\mathcal{H}_1$ & $\mathcal{H}_2$ \\
\midrule
Topological thresholding & 84.26 & 15.12 & 846.92 & 50.03 \\
Min. frequency 2 & 8.47 & 0.77 & 56.55 & 3.37 \\
Min. frequency 5 & 11.03 & 0.87 & 74.10 & 3.77 \\
Frequency 10--150,000 & 10.61 & 0.81 & 69.80 & 3.45 \\
Edge weight $\geq$ 2 & 10.80 & 0.46 & 66.13 & 1.52 \\
Edge weight $\geq$ 5 & 12.39 & 0.37 & 62.68 & 1.11 \\
Disparity & 13.72 & 0.45 & 72.43 & 1.50 \\
LANS & 20.08 & 1.01 & 166.07 & 5.20 \\
Marginal likelihood & 21.90 & 1.37 & 201.46 & 5.80 \\
Hypergeometric & 17.11 & 1.97 & 149.22 & 10.23 \\
Noise-corrected & 14.01 & 5.75 & 110.45 & 34.57 \\
\bottomrule
\end{tabular}}
\par\smallskip\begin{minipage}{\textwidth}\footnotesize \emph{Notes:} Feature densities are counts per 1,000 retained edges ($\mathcal{H}_1$) or triangles ($\mathcal{H}_2$). Total finite persistence is the summed lifetime in years per 1,000 edges or triangles. Counts and sums include positive finite intervals only. Node-based methods retain qualifying isolated vertices; edge-based methods retain only endpoints of surviving edges. Topological thresholding uses constraint percentiles $(50,25)$; statistical thresholds use $\alpha=0.05$, with $z=\Phi^{-1}(0.95)$ for noise-corrected thresholding. All values use the definitions in Appendix~\ref{app:comparison_methods}.
\end{minipage}
\end{table}

\clearpage
\begingroup\setlength{\tabcolsep}{4pt}\renewcommand{\arraystretch}{1.15}
\begin{longtable}{lrrrrr}
\caption{Cross-field comparisons with individual alternatives.}\label{tab:comparison_cross_field}\\
\multicolumn{6}{p{\dimexpr\textwidth-2\tabcolsep\relax}}{\footnotesize Ratios divide the topological-thresholding value by the alternative's value. Wins indicate a higher topological-thresholding value. $n_R$ counts fields with positive values for both methods; $n_P$ counts fields with defined values for both methods, including zeros.}\\[5pt]
\toprule
Alternative & Median & 25th--75th & Win/tie/ & $n_R$ & $n_P$ \\
 & ratio & percentiles & loss & & \\
\midrule
\endfirsthead
\multicolumn{6}{l}{\textbf{Table \thetable\ (continued)}}\\[4pt]
\multicolumn{6}{p{\dimexpr\textwidth-2\tabcolsep\relax}}{\footnotesize Ratios divide the topological-thresholding value by the alternative's value. Wins indicate a higher topological-thresholding value. $n_R$ counts fields with positive values for both methods; $n_P$ counts fields with defined values for both methods, including zeros.}\\[5pt]
\toprule
Alternative & Median & 25th--75th & Win/tie/ & $n_R$ & $n_P$ \\
 & ratio & percentiles & loss & & \\
\midrule
\endhead
\bottomrule
\endfoot
\bottomrule
\endlastfoot
\multicolumn{6}{l}{\textbf{(a) Finite $\mathcal{H}_1$ features per 1,000 edges}}\\[3pt]
Min. frequency 2 & 2.35 & 1.03--5.48 & 109/26/7 & 141 & 142 \\
Min. frequency 5 & 1.48 & 0.54--3.80 & 86/1/55 & 141 & 142 \\
Frequency 10--150,000 & 1.37 & 0.43--3.50 & 83/1/58 & 140 & 142 \\
Edge weight $\geq$ 2 & 2.37 & 1.21--4.27 & 119/1/22 & 138 & 142 \\
Edge weight $\geq$ 5 & 2.78 & 1.43--4.49 & 125/1/16 & 121 & 142 \\
Disparity & 2.78 & 1.41--4.64 & 122/1/19 & 120 & 142 \\
LANS & 2.02 & 1.33--3.86 & 128/1/13 & 140 & 142 \\
Marginal likelihood & 2.41 & 1.49--3.75 & 122/1/19 & 140 & 142 \\
Hypergeometric & 2.01 & 1.27--4.09 & 128/1/13 & 141 & 142 \\
Noise-corrected & 1.54 & 1.05--4.31 & 116/1/25 & 141 & 142 \\
\addlinespace[10pt]
\multicolumn{6}{l}{\textbf{(b) Finite $\mathcal{H}_2$ features per 1,000 triangles}}\\[3pt]
Min. frequency 2 & 2.67 & 1.00--7.22 & 98/27/17 & 137 & 142 \\
Min. frequency 5 & 1.96 & 0.51--6.41 & 85/5/52 & 135 & 142 \\
Frequency 10--150,000 & 1.87 & 0.43--6.75 & 85/5/52 & 133 & 142 \\
Edge weight $\geq$ 2 & 4.99 & 2.44--13.30 & 130/5/7 & 117 & 142 \\
Edge weight $\geq$ 5 & 8.28 & 4.19--19.17 & 132/4/5 & 87 & 141 \\
Disparity & 9.37 & 3.71--15.84 & 133/3/4 & 91 & 140 \\
LANS & 6.38 & 3.55--10.34 & 135/5/2 & 125 & 142 \\
Marginal likelihood & 10.29 & 5.50--20.98 & 136/5/1 & 117 & 142 \\
Hypergeometric & 4.59 & 2.70--7.31 & 131/5/6 & 130 & 142 \\
Noise-corrected & 1.65 & 1.20--2.66 & 112/3/27 & 133 & 142 \\
\newpage
\multicolumn{6}{l}{\textbf{(c) Total finite $\mathcal{H}_1$ persistence per 1,000 edges}}\\[3pt]
Min. frequency 2 & 2.35 & 1.06--5.80 & 107/26/9 & 141 & 142 \\
Min. frequency 5 & 1.49 & 0.52--4.18 & 85/1/56 & 141 & 142 \\
Frequency 10--150,000 & 1.31 & 0.42--3.85 & 84/1/57 & 140 & 142 \\
Edge weight $\geq$ 2 & 2.72 & 1.33--5.54 & 122/1/19 & 138 & 142 \\
Edge weight $\geq$ 5 & 3.82 & 1.67--7.30 & 127/1/14 & 121 & 142 \\
Disparity & 4.06 & 1.79--7.30 & 127/1/14 & 120 & 142 \\
LANS & 2.11 & 1.32--4.38 & 126/1/15 & 140 & 142 \\
Marginal likelihood & 2.69 & 1.51--3.90 & 121/1/20 & 140 & 142 \\
Hypergeometric & 2.39 & 1.22--4.42 & 123/1/18 & 141 & 142 \\
Noise-corrected & 1.52 & 1.01--4.47 & 106/1/35 & 141 & 142 \\
\addlinespace[10pt]
\multicolumn{6}{l}{\textbf{(d) Total finite $\mathcal{H}_2$ persistence per 1,000 triangles}}\\[3pt]
Min. frequency 2 & 2.63 & 1.00--6.83 & 97/27/18 & 137 & 142 \\
Min. frequency 5 & 1.90 & 0.51--5.94 & 84/5/53 & 135 & 142 \\
Frequency 10--150,000 & 1.78 & 0.43--6.79 & 84/5/53 & 133 & 142 \\
Edge weight $\geq$ 2 & 4.95 & 2.46--14.92 & 129/5/8 & 117 & 142 \\
Edge weight $\geq$ 5 & 10.85 & 4.59--22.45 & 132/4/5 & 87 & 141 \\
Disparity & 11.15 & 3.48--20.55 & 133/3/4 & 91 & 140 \\
LANS & 5.85 & 3.22--9.63 & 132/5/5 & 125 & 142 \\
Marginal likelihood & 9.38 & 4.31--21.39 & 134/5/3 & 117 & 142 \\
Hypergeometric & 4.00 & 2.28--6.63 & 130/5/7 & 130 & 142 \\
Noise-corrected & 1.40 & 0.97--2.30 & 100/3/39 & 133 & 142 \\
\end{longtable}\endgroup
\noindent\begin{minipage}{\textwidth}\footnotesize \emph{Notes:} Ratios divide the topological-thresholding value by the named alternative's value within each field. Medians and quartiles are calculated on the logarithmic scale and transformed back; ratios above one favor topological thresholding. $n_R$ counts pairs with two positive values, while $n_P$ counts pairs with two defined values, including zeros. Win/tie/loss counts use the latter sample and the numerical tie tolerance in Appendix~\ref{app:comparison_methods}. Undefined measures and unavailable measurements are excluded from the corresponding pair. The underlying persistence measures in panels (c,d) are in years per 1,000 edges or triangles; their ratios are dimensionless. Topological thresholding uses constraint percentiles $(50,25)$ in every field, and the alternative settings match Table~\ref{tab:comparison_applied_math}.\par\end{minipage}\par

\end{appendices}


\bibliography{sn-bibliography}{}

\end{document}